\documentclass[sigconf,nonacm]{acmart}

\usepackage{amsthm}
\usepackage{array}
\DeclareMathOperator*{\expec}{{\mathbb{E}}} %Expectation
\DeclareMathOperator*{\prob}{\mathbb{P}} %Probability
\DeclareMathOperator*{\var}{Var}	%Variance
\usepackage{dsfont} 
\usepackage{paralist} %For compact enumerations
\usepackage[multiple]{footmisc} %for multiple footnotes to be seperated by a comma
\usepackage{color}

\newtheorem{theorem}{Theorem}

\usepackage[ruled,vlined]{algorithm2e}

\usepackage{enumerate} 

\usepackage{cleveref}
\crefname{equation}{Eq.}{Eqs.}

\usepackage{graphicx}
\usepackage[outdir=./]{epstopdf}
\usepackage{caption}
\usepackage{subcaption}

\newcommand{\degree}{d} % degree function
\newcommand{\avgdegree}{\bar{k}} % average-degree 
\newcommand{\DegDist}{P} % degree-distribution
\newcommand{\conditionalDegDist}{P} % conditional degree-distribution
\newcommand{\conditionalSharingDist}{P_s} % conditional sharing-distribution
\newcommand{\unconditionalSharinProb}{p_s} % unconditional sharing probability
\newcommand{\AssortativityCoefficient}{r} % assortativity coefficient
\newcommand{\DegreeSharingCorrelationCoefficient}{\rho} % degree-sharing correlation coefficient
\newcommand{\shareInfo}{s} %sharing function
\newcommand{\SetOfSharers}{S} %sharing function
\newcommand{\exposureInfo}{f} %exposure function
\newcommand{\neighborset}{\mathcal{N}} %neighbor set
\newcommand{\trueparameter}{\bar{f}} %true exposure function
\newcommand{\stepsize}{\epsilon} %time variable
\newcommand{\exponent}{\alpha} %time variable

\newcommand{\numsamples}{n} %number of samples
\newcommand{\VanillaEstimate}{\hat{f}_{\textrm{Vl}}} %vanilla estimate
\newcommand{\FPEstimate}{\hat{f}_{\textrm{FP}}} %Friendship paradox-based estimate

\newcommand{\DirecteNetwrkRandomNode}{\mathcal{X}} % random node
\newcommand{\DirecteNetwrkRandomFriend}{\mathcal{Y}} % random friend
\newcommand{\DirecteNetwrkRandomFollower}{\mathcal{Z}} % random follower
\newcommand{\FriendbasedEstimate}{\hat{f}_\textrm{Fr}} % Friend based estimate
\newcommand{\FollowerbasedEstimate}{\hat{f}_\textrm{Fo}} % Follower based estimate
\newcommand{\outdegree}{d_{o}} % out-degree function
\newcommand{\indegree}{d_{i}} % in-degree function
\AtBeginDocument{%
  \providecommand\BibTeX{{%
    \normalfont B\kern-0.5em{\scshape i\kern-0.25em b}\kern-0.8em\TeX}}}

\setcopyright{acmcopyright}
\copyrightyear{2022}
\acmYear{2022}
\acmDOI{10.1145/TBD}

\acmConference[In Submission]{KDD}{2022}{KDD}
\acmBooktitle{In Submission, KDD 2022}
\acmPrice{15.00}
\acmISBN{978-1-4503-TBD}

\begin{document}

\title{Estimating Exposure to Information on Social Networks}

\author{Buddhika Nettasinghe}
\email{dwn26@cornell.edu}
\affiliation{%
  \institution{Cornell University}
  \city{}
  \state{}
  \country{}
  \postcode{}
}

\author{Kowe Kadoma}
\email{kk696@cornell.edu}
\affiliation{%
  \institution{Cornell Tech}
  \city{}
  \state{}
  \country{}
}

\author{Mor Naaman}
\email{mor.naaman@cornell.edu}
\affiliation{%
  \institution{Cornell Tech}
  \city{}
  \state{}
  \country{}
}

\author{Vikram Krishnamurthy}
\email{vikramk@cornell.edu}
\affiliation{%
  \institution{Cornell University}
  \city{}
  \state{}
  \country{}
  \postcode{}
}

% \affiliation{%

% \affiliation{%

% \affiliation{%

% \affiliation{%

\renewcommand{\shortauthors}{Nettasinghe and Kadoma, et al.}

\begin{abstract}
This paper 
considers
% presents and defines 
the problem of estimating exposure to information in a social network. Given a piece of information~(e.g.,~a URL of a news article on Facebook, a hashtag on Twitter), our aim is to find the fraction of people on the network who have been exposed to it. 
The exact value of exposure to a piece of information is determined by two features: the structure of the underlying social network and the set of people who shared the piece of information. 
Often, both features are not publicly available~(i.e.,~access to the two features is limited only to the internal administrators of the platform) and difficult to be estimated from data. As a solution, we propose two methods to estimate the exposure to a piece of information in an unbiased manner: a vanilla method which is based on sampling the network uniformly and a method which non-uniformly samples the network motivated by the Friendship Paradox. 
We provide theoretical results which characterize the conditions~(in terms of properties of the network and the piece of information) under which one method outperforms the other.
% \bnedit{In particular, we find that the friendship paradox-based estimator has a smaller variance when the  network assortativity coefficient and the correlation coefficient between sharing and degree have the same sign in undirected networks. The disparity in variance between the two methods is larger when the piece of information is less widely shared or the network has a heavy-tailed degree distribution. }
% \mn{In general I prefer putting the result straight in the abstract: e.g. "We provide theoretical results that show that the method based on the Friendship Paradox works better on high-assortatitivity..." (instead of the previous sentence). But OK to leave as is.}
Further, we outline extensions of the proposed methods to dynamic information cascades (where the exposure needs to be tracked in real-time). 
We demonstrate the practical feasibility of the proposed methods via experiments on multiple synthetic and real-world datasets.

\end{abstract}





\maketitle

\section{Introduction}
\label{sec:intro}

%People are exposed to information on social networks via their friends. 
Online social networks are an important mechanism through which people are exposed to information. Estimating the total number of people who are exposed by their friends to a piece of information\footnote{ A ``piece of information" refers to any uniquely identifiable message~(e.g.,~a URL, a hashtag) that is shared by the users on a social network with their contacts.} on an online social network (e.g., the URL of an article on Facebook, a hashtag on Twitter, etc.) is an important problem with key societal implications. Such 
estimates of exposure
% information 
can, for example, help researchers and the public track the prevalence and reach of election misinformation~\cite{Dwoskin_2020, allcott2017social} or improve the public health response to the Coronavirus~\cite{simpson2020fighting, tasnim2020impact}.

To measure the exposure to a piece of information on a network, one needs access to two features:
the set of people who shared the piece of information, and the structure of the underlying social network. 
% These two \bnedit{parameters}
% resources 
% are often 
% unavailable 
% \bnedit{unknown} to researchers and the public.
Of these two features,
% factors, 
the structure of the underlying social network is often not publicly available and, fully or partially estimating it from data is not a practically feasible task due to the networks' massive size~(e.g.,~billions of nodes and edges in Facebook), constantly evolving nature~\cite{leskovec2008microscopic,leskovec2005graphs}, and limits placed by corporations on data collection. 
Similarly, the set of people who shared the piece of information is often also not publicly known and difficult to estimate from data since it evolves as the piece of information spreads through the social network in the form of an information cascade e.g.,~a URL of a news article that is being shared on Facebook.
% \bn{add the example of a URL on Facebook.} \mn{not sure what the last part of the sentence is arguing.} \bn{Changed it to (hopefully) better convey.} 
As a result, calculating exposure to a piece of information on a social network in a data-driven manner remains a challenging task. 

This state of affairs is unfortunate as efficient (in terms of computation and resources) and accurate (in terms of statistical properties) estimates of exposure to information can provide two important benefits: % main perspectives:
\begin{itemize}[]
\item \emph{1.~Analysis perspective: } %Estimating the exposure to information and its evolution over time is useful 
to identify the pieces of information that are most widely consumed and thus, shed light on how the information consumption patterns affect the outcomes of various high-stake events such as elections, COVID vaccine acceptance, etc.~{\cite{allcott2017social,tasnim2020impact}}. 
% \mn{I switched the order, moved this below the other (more important) contribution; maybe this sentence can be removed altogether as it's not clear:} 
% \bn{Changed to make more precise.} 
% \bnedit{Further, being able to gather data on exposure to information in real-time on social network platforms will also make research experiments on social media, news consumption patterns, etc. easier and more accurate.} 
% \bn{Mention that we can achieve data collection in real time instead of the modified sentence.}
% Further, better exposure data will help researchers understand the relations between various sources and pieces of information in the information ecosystem. %(e.g., using the empirical covariance between them). In addition, they also 

\vspace{0.1cm}
\item \emph{2.~Intervention perspective: } %Being able 
to prioritize various pieces of information in the content moderation and fact checking process of a social media platform. Such interventions could be useful to prevent large numbers of users from being exposed to harmful or misleading information. 
%in scale social media platforms by real-time identification and fact-checking of the trending information.
\end{itemize}

Surprisingly, the problem of estimating exposure to information has received relatively little attention in the literature despite its importance. Formally, this problem can be stated as follows (in the context of an undirected social network such as Facebook).

% \mn{I changed person to "node" below; let's use "user" (when more informal) or "node" (for formal defs) because it's not always a person.}
\vspace{-0.1cm}
\begin{center}
	\fbox{\begin{minipage}{0.98\columnwidth}
			\textbf{Problem of estimating exposure to information:}  Consider an undirected social network  ${G = (V,E)}$ and, let ${\shareInfo(v) = 1}$ if the node $v \in V$ shared~(with the set of their neighbors $\neighborset(v) \subset V$) a piece of information and $\shareInfo(v) = 0$ otherwise. Assuming the graph~$G$ and the sharing function $\shareInfo:V \rightarrow \{0,1\}$ are unknown, estimate the fraction of nodes exposed to the piece of information, 
			 \begin{equation} \label{eq:trueparameter}
			\trueparameter = \frac{ \vert\{v\in V: \exposureInfo(v) = 1\}\vert  }{\vert V\vert} 
			\end{equation} where, $\exposureInfo(v)=1$ if the node $v\in V$ has been exposed to the piece of information by one of their neighbors and $\exposureInfo(v)=0$ otherwise i.e.,~$\exposureInfo(v) = \mathds{1}_{\{\exists\, u\, \in\, \neighborset(v) \text{ such that } \shareInfo(u) = 1\}}(v)$.
		\end{minipage}	
	}
\end{center}

\vspace{0.1cm}
In the above formulation, the value $\shareInfo(v) \in \{0,1\}$ indicates whether the user (i.e., node) $v \in V$ shared the piece of information in concern and $\exposureInfo(v) \in \{0,1\}$ indicates whether the user $v \in V$ has been exposed to it by one of their neighbours. Thus, the parameter of interest $\trueparameter$ denotes the average exposure to the piece of information in the social network~i.e.,~$\trueparameter = \expec\{\exposureInfo(X)\}$ where $X$ denotes a uniformly sampled node from the set of all nodes $V$. In this setting, we are tasked with devising a method to estimate the average exposure~$\trueparameter$. Since the sharing function $\shareInfo:V \rightarrow \{0,1\}$ and the network~$G = (V,E)$ are both unknown, the exposure function $\exposureInfo:V \rightarrow \{0,1\}$ is also unknown~(as it depends on $\shareInfo$ and $G$). However, the exposure $\exposureInfo(v_i) \in \{0,1\}$ of a small number of sampled nodes $v_i, i=1,2,\dots, \numsamples$ (where $\numsamples \ll |V|$) can often be found by looking at whether at least one neighbor of $v_i$~(for each $i=1,2,\dots, \numsamples$) shared the piece of information. 

\vspace{0.2cm}
\noindent
{\bf Main results: } 
% The main results of the current paper are as follows.
%\begin{compactnum}
\begin{compactitem}[]
\item (1) We propose two methods for estimating the exposure to a piece of information in an undirected social network~(i.e.,~the problem that is formalized above): a vanilla method based on uniform sampling and a friendship paradox-based method. Both methods are intuitive and practically feasible. 
% \bnedit{Both methods assume that the values $\exposureInfo(v_1), \exposureInfo(v_2),\dots, \exposureInfo(v_\numsamples) \in \{0,1\}$ of a few sampled nodes $v_1, v_2, \dots, v_\numsamples \in V$ can be found even though the function $\exposureInfo:V \rightarrow \{0,1\}$ is not fully known.}
% (since the sharing function $\shareInfo:V \rightarrow \{0,1\}$ and the graph $G = (V,E)$ on which it depends are not known)
% \bnedit{Both methods assume that the function $\exposureInfo:V \rightarrow \{0,1\}$ (which indicates the exposure to information) is not known fully (since the sharing function $\shareInfo:V \rightarrow \{0,1\}$ and the graph $G = (V,E)$ on which it depends are not known) but, $\exposureInfo(v_i)$ for a few sampled nodes $v_1, v_2, \dots, \numsamples$ can be found. }
% \bnedit{Both methods assume that the function $\exposureInfo:V \rightarrow \{0,1\}$ which indicates the exposure to information is not known fully~(since the sharing function $\shareInfo:V \rightarrow \{0,1\}$ and the graph $G = (V,E)$ on which it depends are not known)}
% \mn{Should we specify whether the two factors (networks, set of people sharing) are unknown? Should we include that in the problem definition?}\bn{changed the problem definition}

\vspace{0.1cm}
\item (2) Via theoretical analysis and numerical experiments, we characterize the conditions~(in terms of the properties of the underlying network and the piece of information in concern) under which the vanilla method outperforms the friendship paradox-based method and vice-versa. {These conditions depend only on parameters that are typically known apriori~(e.g.,~whether the network is assortative or disassortative). As such, these characterizing conditions help to choose the most accurate method for estimating exposure to information depending on the context of the problem.}
%~(e.g.,~a piece of fake news that originated with an unidentified person or, a political opinion expressed by a well-known politician).

\vspace{0.1cm}
\item (3) We extend the two proposed methods to the setting of a dynamic information cascade where the piece of information gradually spreads and more people become exposed to it over time. The resulting algorithms iteratively track
% \mnedit{updating the estimate of} %estimating the 
the 
increasing average exposure
% exposure to a piece of information 
in real-time. %Consequently, decision making~(e.g.,~fact-checking, etc.) and experiments~(e.g.,~evaluating how exposure to a particular piece of news affect COVID vaccine adoption) can be done in real-time~(instead of doing it in the form of a postmortem type experiment). 
In addition, we show how the proposed methods can be extended to the context of directed networks~(e.g.,~Twitter).

\vspace{0.1cm}
\item (4) We provide detailed numerical simulations~(based on synthetic data) as well as empirical experiments~(based on real-world data) to illustrate the usefulness and feasibility of the proposed methods under various practical settings.  
%\end{compactenum}
\end{compactitem}

% \mn{Should we just put all the below under a "related work" section starting right here? I think at this point we said all we wanted to say for the intro.}

% \mn{Totally can remove this paragraph.}\vspace{0.2cm}
% \noindent
% {\bf Organization: } 
% % The rest of this paper is organized as follows. 
% Sec.~\ref{sec:related_work} discusses the recent real-world events that motivate the problem as well as the related work from literature. Sec.~\ref{sec:algorithms} presents the two algorithms for estimating exposure to information on a social network (a vanilla method based on uniform sampling and a friendship paradox-based method) as well as their extensions to dynamic information cascades (where the exposure to information needs to be tracked iteratively). 
% % Further, their extensions to the setting of dynamic information cascades (where the exposure to information needs to be tracked recursively) are also provided.  
% Then, Sec.~\ref{sec:theoretical_comparison} theoretically compares the two methods in terms of their statistical properties. Finally, Sec.~\ref{sec:numerical_results} and Sec.~\ref{sec:empirical_results} present numerical and empirical experiments that complement the statistical analysis and highlight the practical feasibility of the proposed methods in various settings. 

\section{Related Work and Preliminaries}
\label{sec:related_work}

Our problem definition and methods are motivated by recent literature and events, which we expand on below. We then present background preliminaries on the main approach that we use in this work, the friendship paradox.

\subsection{Motivation} 
% Although it has received relatively less attention in the literature, 
The need for a principled method for estimating exposure to information in social media has intensified recently. %There are several reasons for this renewed interest in methods for estimating exposure to information.

One reason for this increased interest in exposure estimation
% need for such methods 
is the role that exposure to information on social networks can play in affecting the outcomes of high-stake events that define the course of our society. 
In particular, the question of reach of ``fake news'' on Facebook has become a major public concern following the 2016 US presidential election~\cite{Dwoskin_2020, allcott2017social}. 
Similarly, large scale exposure to false information on social media has complicated the public health response to the Coronavirus~\cite{simpson2020fighting, tasnim2020impact}. 
Both examples highlight the need for tracking and quantifying exposure, for example to prioritize fact-checking of trending coronavirus and election-related information content. 
%\mntrim{Additionally, \cite{juul2021comparing} recently found that there are no significant structural differences between diffusion patterns of false and true news on social networks. Consequently, machine learning-based detection of false news using the diffusion pattern may not be feasible in such settings and therefore, rapid identification and fact-checking remains the only viable solution to reduce fake news on social media.}

% \mnedit
{It has also become clear that the platforms cannot be trusted to reliably provide this information, in real-time or retroactively~\cite{Ghaffary_2021,Edelman_2021}.}
%Secondly, the dwindling trust of the society towards the information exposure statistics released by social media companies has called for methods that can independently estimate the exposure to various sources of information. 
For example, Facebook recently acknowledged serious problems in the data provided to academic researchers in 2020~\cite{Alba_2021, Timberg_2021},  
%For example, Facebook recently acknowledged that the data provided to academic researchers in 2020 had serious inaccuracies and thus rendered the findings of many academic research papers based on that data unreliable~\cite{Alba_2021, Timberg_2021}. 
%In addition, it has been reported that Facebook also 
and reportedly shelved ``most-viewed pages'' reports when those conflicted with the company's publicity goals~\cite{AlbaMac_2021}. 
%shelved an earlier report on its most-viewed articles due to internal concerns about the possible negative public response to it and, later released a report which is considerably different from the one that was going to be released originally~\cite{AlbaMac_2021}. 
Such incidents have given rise to the need of methods which can accurately estimate the exposure to various pieces of information independently without the involvement of the social media companies. 
%The need for such methods is further bolstered by the reluctance of many social media companies to release enough information that yield a complete picture about implications of exposure to various sources of information~\cite{Ghaffary_2021,Edelman_2021}.

Researchers looking to estimate exposure to information in social networks are largely limited to survey-based methods. 
In such methods, a question is presented to a set of respondents to gather information on the frequency and pattern of their social media usage~\cite{goldman2020debating}. 
For example, researchers had used a post-election online survey to assess how exposure to fake news affected the 2016 US election~\cite{allcott2017social}.
% \mnedit
{Other research had used panels of users who provided access to their web traffic to assess such exposure~\cite{guess2020exposure}. 
Closer to our baseline method here, researchers had used a panel of Twitter users to track their exposure to specific "fake news" URLs and domains shared by people they follow, but did not offer network-wide measures~\cite{grinberg2019fake}.}

%Despite being easy to implement, survey-based methods have several limitations. First, they cannot be implemented in real-time and therefore, can only be used for post-event type studies. Consequently, they are not suitable for purposes such as identifying trending information for independent fact-checking etc. Second, as respondents self-report their usage frequencies and patterns, the outcomes of surveys are prone to over- and under- reporting errors as well as human cognitive biases~\cite{konitzer2021comparing}. Consequently, the estimates of exposure to information based on surveys alone may not have rigorous theoretical guarantees on the accuracy. 
As opposed to the post-event survey-based approach, the social network sampling-based methods proposed in this work can be implemented in real-time (to track the progression of exposure as a piece of information spreads over time).
Further, these methods yield unbiased estimates of the exposure to information across the entire social network. 
In addition, these techniques can be implemented in a practically feasible manner without the full knowledge of the network~(e.g.,~via a random walk) as well as the set of people who shared the piece of information. 

\subsection{The Friendship Paradox}

%One estimator for the average exposure to information that we propose is based on uniform sampling of nodes while the other is based on sampling random friends~(i.e.,~random ends of random edges). 
%The latter 
Our work here
is motivated by the graph theoretic consequence named \emph{friendship paradox} which states ``on average, the number of friends of a random friend is always greater than or equal to the number of friends of a random individual". Formally:
\begin{theorem}[Friendship paradox~\cite{feld1991your}]
\label{th:friendship_paradox}
Consider an undirected graph $G = (V, E)$. Let $X$ be a node sampled uniformly from $V$ and $Y$ be a uniformly sampled end-node from a uniformly sampled edge $e \in E$. Then,
\begin{equation}
    \expec\{\degree(Y)\} \geq \expec\{\degree(X)\},
\end{equation}
where $\degree(X)$ and $\degree(Y)$ denote the degrees of $X$ and $Y$, respectively. 
\end{theorem}

In Theorem~\ref{th:friendship_paradox}, the random variable $Y$ is called a random friend. This is because it is a random person from a uniformly sampled pair of friends.\footnote{A random friend $Y$ on an undirected graph $G = (V, E)$ has the distribution $\prob\{Y = v\} \propto \degree(v)$ for all $v \in V$. In other words, a random friend is a node sampled with a probability proportional to their degrees.} The intuition behind the friendship paradox (Theorem~\ref{th:friendship_paradox}) is that individuals with large number of friends~(i.e.,~high-degree nodes) appear as the friends of many others. Therefore, a random end of a random link~(i.e.,~the random variable~$Y$) is more likely to yield a high degree node than a uniformly sampled node~(i.e.,~the random variable~$X$). Consequently, sampling random friends~(i.e.,~$Y$) allows us to reach high-degree nodes in the network without the full knowledge of the network. 

% \mn{Can't just say "this" without a reference. "This idea",or "This technique"... always have a reference. Same with "They" or "it". Please pay attention to it throughout the paper.} 
% This \mnedit{insight} 
The friendship paradox
has been exploited in many statistical estimation methods
% proposed in literature, 
% for example,
% Such uses of the friendship paradox included 
% Some examples of such uses of the friendship paradox include
% methods 
e.g.,~to reduce the variance in survey based polling methods~\cite{nettasinghe2019your}, to efficiently estimate power-law degree distributions~\cite{nettasinghe2021maximum, eom2015tail}, and to quickly detect the outbreak of a disease~\cite{garcia2014using}. 
% Compared to the existing methods, the methods that we propose are unbiased (even for finite sample sizes) and works in real-time.
{This paper differs from such existing friendship paradox-based methods in several key ways. First, our focus is on the problem of estimating the fraction of people exposed to a piece of information, which is different from the problems studied 
% \kkedit{[maybe state what those problems are]}
in prior works
(such as polling, estimating degree distributions, detecting disease outbreaks) 
that are based on the friendship paradox. 
Second, we do not claim that the friendship paradox-based method is the better approach for any given setting. Instead, we provide an exact characterization of the conditions where the friendship paradox-based method outperforms the vanilla method based on uniform sampling.
Finally, the methods that we propose are provably unbiased whereas most previously proposed estimators based on the friendship paradox attempt to trade-off the bias and variance in order to achieve a smaller mean squared error.}

%\section{Algorithms to Estimate Exposure to Information}
\section{Algorithmic Approach}
\label{sec:algorithms}

In this section, we present two methods for estimating the average exposure to information: a vanilla method based on uniform sampling and a friendship paradox-based method. 
% We also consider the extensions of these methods to the context of dynamic information cascades. %are also discussed. 
	
\subsection{Vanilla method based on uniform sampling}
\label{subsec:VanillaEstimate}

The vanilla approach for estimating the average exposure $\trueparameter$ works by obtaining a set of random nodes and checking whether these nodes have been exposed to the piece of information via their contacts.

%as follows.
\vspace{0.2cm}
\noindent
\emph{Vanilla method for estimating exposure to information}
\vspace{0.1cm}
\begin{compactitem}[]
\item \textit{Step 1:} Sample $\numsamples$ random nodes $X_1, \dots, X_\numsamples$ uniformly and independently from the set of all nodes $V$.
\vspace{0.1cm}
\item \textit{Step 2:} Use, 
	    \begin{equation}
	        \label{eq:VanillaEstimate}
	        \VanillaEstimate = \frac{\sum_{i = 1}^{n}{\exposureInfo(X_i)}}{\numsamples}
        \end{equation}
        as the estimate of the average exposure $\trueparameter$.
\end{compactitem}
  
  \vspace{0.2cm}
 The vanilla estimator $\VanillaEstimate$ given in \cref{eq:VanillaEstimate} is unbiased~i.e.,~$\expec\{\VanillaEstimate\} = \trueparameter$. However, the vanilla estimator $\VanillaEstimate$ would intuitively yield a larger variance since random nodes are not likely exposed to the piece of information when $\shareInfo(\cdot)$ is a very sparse function (i.e.,~only very few people shared the information). 

\subsection{Friendship paradox-based method}
\label{subsec:FPEstimate}

In order to reduce the variance in the estimate of average exposure, we can exploit the friendship paradox-based sampling~(instead of 
vanilla uniform sampling) as follows:
% In order to reduce the variance in the estimate, instead of 
% \bnedit{sampling uniformly,}
% % using a 
% % \bnedit{uniform}
% % random 
% % sample
% we can exploit the friendship paradox-based sampling as follows:

\vspace{0.2cm}
\noindent
\emph{Friendship paradox-based method for estimating exposure to information}
\begin{compactitem}[]
\vspace{0.1cm}
\item \textit{Step 1:} Sample $n$ random friends $Y_1, \dots, Y_\numsamples$ from the network independently (a random friend $Y_i$ is a random end of a random link~i.e.,~a link is sampled uniformly from the network and one end of that link is taken with an unbiased coin flip).
\vspace{0.1cm}
		
\item \textit{Step 2:} Use,
		\begin{equation}
		\label{eq:FPEstimate}
		\FPEstimate = \frac{\avgdegree}{n}\sum_{i = 1}^{\numsamples}\frac{\exposureInfo(Y_i)}{\degree(Y_i)}
		\end{equation}
		as the estimate of average exposure $\bar{f}$ where, $\degree(v)$ denotes the degree of $v \in V$ and $\avgdegree$ the average degree of the graph $G = (V, E)$.
\end{compactitem}

\vspace{0.1cm}
The friendship paradox-based estimator~$\FPEstimate$ can be viewed as an application of importance sampling in social networks where the samples are generated from a 
different distribution~(i.e.,~random friends~$Y$) than the distribution that is directly related to the
% of direct interest~(i.e.,~random nodes~$X$) in order to reduce the variance of the 
parameter of interest $\trueparameter = \expec\{f(X)\}$~(i.e.,~random nodes~$X$). 
 In particular, random friends are more popular than random nodes in expectation~(according to Theorem~\ref{th:friendship_paradox}) and thus, sampling random friends will lower the variance by 
 accessing
%  incorporating 
 individuals who are more likely to be exposed to the piece of information due to their large popularity~(even when the sharing function $\shareInfo(\cdot)$ is sparse).  
% Since random friends are more popular than random nodes in expectation~(according to Theorem~\ref{th:friendship_paradox}), the friendship paradox-based estimator~$\FPEstimate$ can yield a smaller variance even for a piece of information that has a small sharing~(i.e.,~function $\shareInfo(\cdot)$ is sparse) by accessing the individuals who are more likely to be exposed to it due to their popularity. 
The additional terms $\degree(Y_i), \avgdegree$ that appear in $\FPEstimate$~(compared to $\VanillaEstimate$) correct for the bias resulting from sampling the more popular random friends~$Y_i$ instead of random nodes~$X_i$. 

{The average degree $\avgdegree$ in Eq.~\ref{eq:FPEstimate} is typically a known statistic for most social networks such as Facebook~\cite{ugander2011anatomy}, which makes the implementation of the friendship paradox-based estimator $\FPEstimate$ practically feasible. Further, in situations where the edges cannot be sampled uniformly from the network, the friendship paradox-based estimator $\FPEstimate$ can also be implemented via random walks as described in Appendix~\ref{sec:appendix_extensions}.}

% \bn{With the addition of the new ``Extensions" section, I think we can remove this paragraph} Note that our discussion so far considered undirected networks~(e.g., Facebook). For directed networks (e.g.,~Twitter), the notion of a ``random friend'' that we used so far %does not extend trivially due to the asymmetry of links. 
% \mnedit
% {needs a slight modification.} 
% In particular, we have three methods to sample a directed network: random node, random friend (tail end of a random link) and random follower (source end of a random link). 
% Consequently, we can have three 
% estimates to estimate 
% \bnedit{estimators for}
% the exposure to information in directed networks. These three methods are discussed in detail in Appendix~\ref{sec:appendix_additional_details_and_more_results}. 
% \mnedit
% {In addition, our empirical results in Sec.~\ref{sec:empirical_results} are based on directed networks.}
% \end{remark}

To summarize, Sec.~\ref{sec:algorithms} presented two methods to estimate the average exposure to information based on uniform~(vanilla) and friendship paradox-based sampling. Extensions of the two proposed methods will be discussed in Sec.~\ref{sec:extensions}.

\section{Comparison of Statistical Properties of the two methods}
\label{sec:theoretical_comparison}

In this section, we analyze and compare the statistical properties of the two proposed estimators (the vanilla estimator~$\VanillaEstimate$ given in \cref{eq:VanillaEstimate} and the friendship paradox-based estimator $\FPEstimate$ given in \cref{eq:FPEstimate}). The aim of this analysis is to identify the conditions under which one estimator may be more accurate than the other for estimating the average exposure to information~$\trueparameter$.

The following result~(see Appendix~\ref{subsec:proof_th_Bias_and_Variance} for proof) characterizes the bias and variance of the two exposure estimators.
\begin{theorem}
\label{th:Bias_and_Variance}
Consider the vanilla estimator~$\VanillaEstimate$ given in \cref{eq:VanillaEstimate} and the friendship paradox-based estimator~$\FPEstimate$ given in \cref{eq:FPEstimate}. 
\begin{compactenum}
     \item Both the vanilla estimator~$\VanillaEstimate$ and the friendship paradox-based estimator~$\FPEstimate$ are unbiased estimators of the fraction of people exposed to a piece of information $\trueparameter$ (defined in \cref{eq:trueparameter})~i.e.,~
     \begin{equation}
     \label{eq:bias}
         \expec\left\{\VanillaEstimate\right\} = \expec\left\{\FPEstimate\right\} = \trueparameter.
     \end{equation}
     
     \item The variances of vanilla estimator $\VanillaEstimate$ and the friendship paradox-based estimator~$\FPEstimate$ are,
     \begin{align}
     \begin{split}
     \label{eq:variance}
         \var\left\{\VanillaEstimate\right\} &= \frac{1}{\numsamples}\trueparameter\left(1-\trueparameter\right)\\
        \var\left\{\FPEstimate\right\} &= \frac{1}{\numsamples}\left( \avgdegree\expec\left\{\frac{\exposureInfo(X)}{\degree(X)}\right\} - \trueparameter^2\right)
     \end{split}
     \end{align}
     where $\avgdegree$ is the average degree of the graph.
     \end{compactenum}
\end{theorem}
% \begin{proof}
% See Appendix~\ref{subsec:proof_th_Bias_and_Variance}.
% \end{proof}

As stated in the first part of Theorem~\ref{th:Bias_and_Variance}, both the vanilla estimator~$\VanillaEstimate$ and the friendship paradox-based estimator~$\FPEstimate$ are unbiased estimators of the average exposure $\trueparameter$. Therefore, the method that has the smaller variance in the given setting should be used for estimating the average exposure $\trueparameter$. In order to do this, the rest of this section aims to identify the conditions under which one method outperforms the other in terms of the variance. 

To theoretically compare the variances, we consider the class of undirected Markovian random networks that are completely characterized by their degree distribution~$\DegDist(k)$ (which gives the probability that a uniformly sampled node from the network has degree $k$) and the conditional degree distribution $\conditionalDegDist(k'|k)$ (which gives the conditional probability that an edge from a degree $k$ node connects to a degree $k'$ node). The term ``Markovian" here refers to the fact that all higher-order correlations can be expressed only in terms of the two functions $\DegDist(k)$ and $\conditionalDegDist(k'|k)$. We can derive a joint degree distribution using $\DegDist(k)$ and $\conditionalDegDist(k'|k)$ as
\begin{equation}
    \label{eq:joint_degree_distribution}
    \DegDist\left(k,k'\right) = \frac{k\DegDist(k)}{\avgdegree} \DegDist\left(k'|k\right),
\end{equation}
which gives the probability that a uniformly sampled link connects two nodes with degrees $k$ and $k'$. The correlation coefficient corresponding to this joint degree distribution $\DegDist\left(k,k'\right)$ is called the \emph{assortativity coefficient}, and we denote it with~$\AssortativityCoefficient \in [-1,1]$. Networks for which $\AssortativityCoefficient > 0$ are called assortative networks since high-degree nodes are more likely to be connected to other high-degree nodes and low-degree nodes are more likely to be connected to other low-degree nodes. On the other hand, networks for which $\AssortativityCoefficient < 0$ are called disassortative networks since high-degree nodes and low-degree nodes are more likely to be connected with each other. A detailed description of Markovian random networks and assortativity can be found in~\cite{bogua2003epidemic,boguna2002epidemic}. 

In addition to $\DegDist(k)$ and $\conditionalDegDist\left(k'|k\right)$ which characterize the Markovian random network, we also define 
% another function 
$\conditionalSharingDist(1|k)$ which is the conditional probability that a node with degree $k$ shares the piece of information. 
Consequently, $\conditionalSharingDist(0|k) = 1 - \conditionalSharingDist(1|k)$ is the probability that a node with degree $k$ does not share the piece of information. 
Intuitively, if the $\conditionalSharingDist\left(1|k\right)$ is closer to $1$ for larger~(resp.~smaller) values of $k$, then high-degree~(resp.~low-degree) nodes are more likely to share the piece of information. In particular, if the sharing happens independently of the node popularity (i.e.,~high-degree and low-degree nodes are equally likely to share the piece of information), then $\conditionalSharingDist(1|k)$ would be a constant that does not depend on degree $k$. 
Such relations (between sharing and degree) can be captured using the correlation coefficient between the sharing and the degree, which we refer to as the \emph{degree-sharing correlation coefficient} and denote by $\DegreeSharingCorrelationCoefficient \in [-1,1]$.

The following result (see Appendix~\ref{subsec:proof_th_variance_comparison_condition} for proof) compares the variances of the two estimators $\FPEstimate, \VanillaEstimate$~(given in \cref{eq:VanillaEstimate} and \cref{eq:FPEstimate}, respectively) in terms of the degree distribution $\DegDist(k)$, the conditional degree distribution $\conditionalDegDist\left(k'|k\right)$ and the conditional sharing probability $\conditionalSharingDist(1|k)$ in the context of Markovian random networks.

\begin{theorem}
\label{th:variance_comparison_condition}
The variance of the friendship paradox-based estimator~$\FPEstimate$ given in \cref{eq:FPEstimate} is less than or equal to the variance of the vanilla estimator~$\VanillaEstimate$ given in \cref{eq:VanillaEstimate} (i.e.,~$\var\{\FPEstimate\} \leq \var\{\VanillaEstimate\}$) if and only if
\begin{equation}
\label{eq:variance_comparison_condition}
\mathbb{E}_{k \sim \DegDist(k)} \left\{ \left(1 - \frac{\avgdegree}{k}\right)\prob\left\{\exposureInfo(X) = 1|\degree(X) = k\right\}\right\} \geq 0,
\end{equation}
where $X$ is a uniformly sampled node from the network, $\mathbb{E}_{k \sim \DegDist(k)}$ denotes the expectation with respect to the degree distribution $\DegDist(k)$, and 
\begin{equation}
    \label{eq:prob_degree_k_exposed}
   \prob\left\{f(X) = 1|\degree(X) = k\right\} = 1 - \left( \sum_{k'}\conditionalDegDist\left(k'|k\right)\conditionalSharingDist\left(0|k'\right)\right)^k. 
\end{equation}
\end{theorem}
% \begin{proof}
% See Appendix~\ref{subsec:proof_th_variance_comparison_condition}.
% \end{proof}

% \vspace{0.1cm}
\noindent
{\bf Discussion of Theorem~\ref{th:variance_comparison_condition}: } Theorem~\ref{th:variance_comparison_condition} yields insights that help identify the settings where one method is more accurate (in terms of variance) compared to the other for estimating the average exposure to information $\trueparameter$. In particular, these insights relate the variance of the methods to important network parameters such as the degree distribution~$\DegDist(k)$, assortativity coefficient~$\AssortativityCoefficient$ and the degree-sharing correlation coefficient~$\DegreeSharingCorrelationCoefficient$ as we discuss in detail below. 

{\it 1.~Choosing the best method based on assortativity coefficient $\AssortativityCoefficient$ and degree-sharing correlation coefficient $\DegreeSharingCorrelationCoefficient$: } Due to the term $\left(1 - {\avgdegree}/{k}\right)$, the condition \cref{eq:variance_comparison_condition} is more likely to be satisfied 
%(indicating that the friendship paradox based estimate $\FPEstimate$ has a smaller variance compared to the vanilla estimate $\VanillaEstimate$)
%$\var\left\{\VanillaEstimate\right\} \geq \var\left\{\FPEstimate\right\}$ 
when the value $\prob\left\{f(X) = 1|\degree(X) = k\right\}$ is closer to $1$ for larger values of the degree $k$ and the value $\prob\left\{f(X) = 1|\degree(X) = k\right\}$ is closer to $0$ for smaller values of the degree $k$. According to \cref{eq:prob_degree_k_exposed}, this happens when,
\begin{enumerate}[i.]
    \item $\conditionalDegDist\left(k'|k\right)$ is closer to $1$ for $k'\gg \avgdegree \gg k$ and $\conditionalSharingDist\left(1|k'\right)$ is closer to $0$ for $k' \gg \avgdegree$,\\
    or,
    \item $\conditionalDegDist\left(k'|k\right)$ is closer to $1$ for $k, k'\ll \avgdegree$ and $\conditionalSharingDist\left(1|k'\right)$ is closer to $0   $ for $k' \ll \avgdegree$.
\end{enumerate}
Consequently, friendship paradox based estimator $\FPEstimate$ has a smaller variance compared to the vanilla estimator $\VanillaEstimate$ % $\var\left\{\VanillaEstimate\right\} \geq \var\left\{\FPEstimate\right\}$ 
when $\AssortativityCoefficient, \DegreeSharingCorrelationCoefficient>0$ (i.e.,~the network is assortative and the high-degree individuals are more likely to share the piece of information) or $\AssortativityCoefficient, \DegreeSharingCorrelationCoefficient<0$ (i.e.,~the network is disassortative and the low-degree individuals are more likely to share the piece of information). {The arithmetic signs of the assortativity coefficient $\AssortativityCoefficient$ and the degree-sharing correlation coefficient $\DegreeSharingCorrelationCoefficient$ are typically known parameters based on the context. For example, it is known that many social networks are assortative whereas technological networks are disassortative~\cite{newman2002assortative}. Similarly, for pieces of information that get originated from highly popular individuals~(e.g.,~political news, updates on government policy, etc.), the degree-sharing correlation coefficient is typically positive. Hence, this first insight allows us to choose the estimator that best suits the given context based on arithmetic signs of the assortativity coefficient $\AssortativityCoefficient$ and the degree-sharing correlation coefficient $\DegreeSharingCorrelationCoefficient$.}

\vspace{0.1cm}
{\it 2.~When the sharing is independent from the popularity: } If the node degree and the sharing are statistically independent, \cref{eq:prob_degree_k_exposed} yields that $\prob\left\{f(X) = 1|\degree(X)\right\} = 1-\conditionalSharingDist(0)^{k}$, where $\conditionalSharingDist(0)=1-\conditionalSharingDist(1)$ is the probability that any node (independent of the degree $k$) does not share the piece of information. Consequently, when the node degree and the sharing are statistically independent,  the friendship paradox-based estimator $\FPEstimate$ has a smaller variance compared to the vanilla estimator $\VanillaEstimate$ if and only if,
\begin{equation}
    \label{eq:variance_comparison_condition_independent_case}
    \mathbb{E}_{k \sim \DegDist(k)} \left\{ \left(1 - \frac{\avgdegree}{k}\right)\left(1-\conditionalSharingDist(0)^{k}\right)\right\} \geq 0, 
\end{equation}
according to \cref{eq:variance_comparison_condition}. We numerically evaluated the expected value in the left-hand-side of \cref{eq:variance_comparison_condition_independent_case} for two types of degree distributions: a power-law distribution $\DegDist(k) \propto k^{-\exponent}$ (where $\exponent >2$ is the power-law exponent) and an exponential distribution $\DegDist(k) = \frac{1}{\lambda}e^{-\lambda k}$ (where $\lambda >0$ is a constant parameter). Our results 
% \bnedit{(detailed in the Appendix)} \bn{This result is not yet added.} 
suggest that the condition \cref{eq:variance_comparison_condition_independent_case} is not satisfied for all values of the power-law exponent $\exponent>2$, the exponential parameter $\lambda>0$, and the probability $\conditionalSharingDist(0) \in [0,1]$. Therefore, the vanilla estimator $\VanillaEstimate$ produces the theoretically better estimate when the node degree and the sharing are statistically independent~(implying that $\DegreeSharingCorrelationCoefficient = 0$) and the network has either a power-law or an exponential degree distribution. 

\vspace{0.1cm}
{\it 3.~A widely shared piece of information vs. a less widely shared piece of information: } According to \cref{eq:variance_comparison_condition_independent_case}, smaller values of $\conditionalSharingDist(0)$ lead to a bigger disparity in the performance between the friendship paradox-based estimator $\FPEstimate$ and the vanilla estimator $\VanillaEstimate$. This implies that choosing the right method is of particular importance when the fraction of people who share a piece of information is small (compared to the size of the network). An important example for this situation is the starting phase of an information cascade where a piece of information has been less widely shared. Since identifying the correct exposure at this beginning stage is crucial for purposes such as fact-checking before many people are exposed, this further highlights the importance of choosing the right method by using the insight~(1) discussed earlier. 

In summary, Sec.~\ref{sec:theoretical_comparison} theoretically analyzed the bias and variance of the two estimators presented in Sec.~\ref{sec:algorithms} (vanilla estimator $\VanillaEstimate$ and the friendship paradox-based estimator $\FPEstimate$) in terms of properties of the underlying network and the piece of information. Next, we present some extensions of the proposed methods~(in Sec.~\ref{sec:extensions}), and then verify and complement the theoretical insights using numerical experiments~(in Sec.~\ref{sec:numerical_results}).

\section{Extensions of Proposed Methods}
\label{sec:extensions}

In this section, we extend the vanilla estimator~$\VanillaEstimate$ in \cref{eq:VanillaEstimate} and the friendship paradox-based estimator~$\FPEstimate$ \cref{eq:FPEstimate} to directed networks~(Sec.~\ref{subsec:directed_networks}) and dynamic information cascades~(Sec.~\ref{subsec:stochastic_approximation}). 

\subsection{Directed networks} 
\label{subsec:directed_networks}

In a directed network $G = (V,E)$ (e.g.,~Twitter), a link $u\rightarrow v$ (pointing from a node $u\in V$ to $v \in V$) indicates that the node $v$ follows the node $u$~i.e.,~$u$ is the friend and $v$ is the follower. Hence, the out-degree $\outdegree(v)$ and in-degree $\indegree(v)$ of a node $v \in V$ denote the number of followers and friends of $v$, respectively. 
% Analogous to the case of undirected networks, w
We say that a node $v\in V$ in a directed network is exposed to a piece of information~(i.e.,~$\exposureInfo(v)=1$) if at least one friend of $v$ shared it. In this context, our aim is to estimate the average exposure to the piece of information which is denoted by $\trueparameter$.

In order to estimate the average exposure $\trueparameter$, a directed network can be sampled in three different ways: a random node $\DirecteNetwrkRandomNode$~(sampled uniformly from $V$), a random friend~$\DirecteNetwrkRandomFriend$ which is the tail end of a uniformly sampled link~(i.e.,~$\prob(\DirecteNetwrkRandomFriend = v) \propto \outdegree(v)$), a random follower~$\DirecteNetwrkRandomFollower$ which is the source end of a uniformly sampled link~(i.e.,~$\prob(\DirecteNetwrkRandomFollower = v) \propto \indegree(v)$). Consequently, we can construct three estimators of the average exposure $\trueparameter$ as follows:
\begin{equation}
    \VanillaEstimate = \frac{\sum_{i = 1}^{n}{\exposureInfo(\DirecteNetwrkRandomNode_i)}}{\numsamples}, \hspace{0.1cm}
    % \quad \text{where $\DirecteNetwrkRandomNode_i,\dots,\DirecteNetwrkRandomNode_\numsamples$ are iid random nodes},        \label{eq:DirectedVanillaEstimate}\\
	\FriendbasedEstimate = \frac{\avgdegree}{n}\sum_{i = 1}^{\numsamples}\frac{\exposureInfo(\DirecteNetwrkRandomFriend_i)}{\outdegree(\DirecteNetwrkRandomFriend_i)}, \hspace{0.1cm}
% 	\quad \text{(where $\DirecteNetwrkRandomFriend_i,\dots,\DirecteNetwrkRandomFriend_\numsamples$ are iid random friends)}, \label{eq:DirectedFriendEstimate}   \\ 
	\FollowerbasedEstimate = \frac{\avgdegree}{n}\sum_{i = 1}^{\numsamples}\frac{\exposureInfo(\DirecteNetwrkRandomFollower_i)}{\indegree(\DirecteNetwrkRandomFollower_i)} 
% 	\quad \text{(where $\DirecteNetwrkRandomFollower_i,\dots,\DirecteNetwrkRandomFollower_\numsamples$ are iid random followers)}, 
% \label{eq:DirectedFollowerEstimate}    
\label{eq:DirectedEstimates}
\end{equation}
where $\avgdegree$ corresponds to the average in-degree $\expec\{\indegree(\DirecteNetwrkRandomNode)\}$ (which is also same as the average out-degree $\expec\{\outdegree(\DirecteNetwrkRandomNode)\}$) and indices $i = 1, \dots, \numsamples$ denote iid samples of each sampling method. 
The three estimators given in \cref{eq:DirectedEstimates} are motivated by the four versions of the friendship paradox that can exist on directed networks~\cite{alipourfard2020friendship,higham2019centrality}. In particular, one version says that a random follower on average has more friends than a random node~(i.e.,~$\expec\{\indegree(\DirecteNetwrkRandomFollower)\} \geq \expec\{\indegree(\DirecteNetwrkRandomNode)\}$), implying that random followers ($\DirecteNetwrkRandomFollower_i$) are more likely to be exposed to a piece of information. Hence, $\FollowerbasedEstimate$ in \cref{eq:DirectedEstimates} can reduce the variance by incorporating more exposed individuals into the sample.

\subsection{Dynamic information cascades}
\label{subsec:stochastic_approximation}

The vanilla estimator $\VanillaEstimate$ in~\cref{eq:VanillaEstimate} and the friendship paradox-based estimator $\FPEstimate$ in~\cref{eq:FPEstimate}
assume that
% can be used to estimate exposure to information when 
the function $\shareInfo(\cdot)$ indicating the set of people who have shared the piece of information 
is static.
%remains invariant over time i.e.,~the piece of information in concern does not spread over time any more. 
However, the set of people who have shared a piece of information typically grows over time as it gets reshared and reposted by the users who were exposed it, leading to an information cascades.
% ~\cite{cheng2014can}.
% in the form of an information cascade.
%e.g.,~the URL of a news article on Facebook, a hashtag on Twitter. In such dynamic information cascades, the set of people who share the piece of information~(defined by the function $\shareInfo(\cdot)$) grows over time and consequently, the set of people who are exposed to the piece of information (defined by the function $\exposureInfo(\cdot)$) also increase over time. 
%To deal with this dynamic setting, 
This subsection extends the %discusses how the 
vanilla estimator $\VanillaEstimate$ and the friendship paradox-based estimator $\FPEstimate$ %can be modified 
to track the increasing average exposure to such information cascades in real-time. The key idea is to use a stochastic approximation algorithm with a constant step-size. % as we discuss in detail next. 

To simplify the notation, let us assume that only one sample can be collected at each time instant and there are no samples at time~$0$, %Hence, the number of samples at any time instant would be equal to that time instant, 
allowing us to use the same variable $\numsamples$ for discrete time and the number of samples available. 
Further, let the vanilla and friendship paradox-based estimators at time $\numsamples$ be denoted by $\VanillaEstimate^{(\numsamples)}$ and $\FPEstimate^{(\numsamples)}$ respectively. Then, note that,
\begin{align}
\begin{split}
    \label{eq:SA_Recursions}
    \VanillaEstimate^{(\numsamples)} &=  \VanillaEstimate^{(\numsamples-1)} + \frac{1}{\numsamples} \left( \exposureInfo\left(X_{\numsamples}\right) - \VanillaEstimate^{(\numsamples-1)}\right)\\
    \FPEstimate^{(\numsamples)} &=  \FPEstimate^{(\numsamples-1)} + \frac{1}{\numsamples} \left( \avgdegree\frac{\exposureInfo\left( Y_{\numsamples}\right)}{\degree\left(Y_{\numsamples}\right) } - \FPEstimate^{(\numsamples-1)}\right)
\end{split}
\end{align}
where, $X_{\numsamples}, Y_{\numsamples}$ denote a random node and a random friend at time $\numsamples$, respectively. 
% The first recursion in \cref{eq:SA_Recursions} can be obtained by observing that the difference between vanilla estimate with $\numsamples$ samples and $\numsamples-1$ samples (i.e.,~$\VanillaEstimate^{(\numsamples)} -  \VanillaEstimate^{(\numsamples-1)} = \frac{\sum_{i = 1}^{\numsamples}{\exposureInfo(X_i)}}{\numsamples} - \frac{\sum_{i = 1}^{\numsamples-1}{\exposureInfo(X_i)}}{\numsamples-1}$) is equal to the update term $\frac{1}{\numsamples} \left( \exposureInfo\left(X_{\numsamples}\right) - \VanillaEstimate^{(\numsamples-1)}\right)$. The second recursion in \cref{eq:SA_Recursions} also follow from a similar argument for the friendship paradox based-estimate. Thus, the recursions in \cref{eq:SA_Recursions} take the intuitive form that the new estimate (from any of the two methods: vanilla or friendship paradox-based) with $\numsamples$ samples is equal to the previous estimate with $\numsamples-1$ samples plus an update term. The additive update term is simply the scaled (by $1/\numsamples$ factor) difference between the estimate computed with the single new sample collected at time $\numsamples$ and the value of the recursion at the previous time instant. Also, note 
The recursions in \cref{eq:SA_Recursions} are obtained under the assumption that the average exposure $\trueparameter$ is time-invariant and therefore, the update term decays with time (due to the decreasing step-size $1/\numsamples$) and converges to zero. Intuitively, this means that one new sample would not make a significant difference to an estimate (of a time-invariant parameter) derived with a relatively large number of samples~(i.e.,~$\numsamples \gg 1$). In particular, it can be shown that the recursions in \cref{eq:SA_Recursions} converge to the average exposure $\trueparameter$ with probability $1$ under mild conditions.

% Let us now turn to the case where the average exposure to information evolves over time as a consequence of the piece of information spreading in the form of an information cascade. Let the average exposure at time $\numsamples = 1,2,...$ be denoted by $\trueparameter^{(\numsamples)}$. Since the average exposure $\trueparameter^{(\numsamples)}$ is now evolving over time, the estimates need to be updated at each time instant (irrespective of how large the number of samples $\numsamples$ is). 

However, the decreasing step-size $1/\numsamples$ in \cref{eq:SA_Recursions} is not suitable when the average exposure is evolving over time~(denoted by~$\trueparameter^{(\numsamples)}$). This is because the decreasing step-size $1/\numsamples$ 
% keep updating less and less over time and 
will stop updating eventually even though the average exposure $\trueparameter^{(\numsamples)}$ will keep changing. As a solution, the decreasing step-size in \cref{eq:SA_Recursions} can be replaced with a constant step-size $\stepsize > 0$ for the case of time evolving average exposure $\trueparameter^{(\numsamples)}$. Then, the new recursive methods for tracking the time evolving average exposure $\trueparameter^{(\numsamples)}$ using the vanilla and friendship paradox-based methods will be as follows:
\begin{align}
	    \VanillaEstimate^{(\numsamples)} &=  \VanillaEstimate^{(\numsamples-1)} + \stepsize \left( \exposureInfo\left(X_{\numsamples}\right) - \VanillaEstimate^{(\numsamples-1)}\right) \label{eq:vanilla_SA}\\
	    \FPEstimate^{(\numsamples)} &=  \FPEstimate^{(\numsamples-1)} + \stepsize \left( \avgdegree\frac{\exposureInfo\left( Y_{\numsamples}\right)}{\degree\left(Y_{\numsamples}\right) } - \FPEstimate^{(\numsamples-1)}\right). \label{eq:FP_SA}
\end{align}

The above two methods can track the progression of average exposure $\trueparameter^{(\numsamples)}$ when it is evolving on a slower time scale compared to the collection of samples. In other words, $\trueparameter^{(\numsamples)}$ is assumed to remain approximately constant for every $c > 1$ samples being collected; if $c \approx 1$~(resp.~$c \gg 1$), we say the piece of information is spreading rapidly~(resp.~slowly). 
The value of the step-size parameter $\stepsize>0$ in \cref{eq:vanilla_SA} and \cref{eq:FP_SA} determines the effect of the update at each time. 
In particular, the value of $\stepsize$ should be relatively large~(resp.~small) to track the the average exposure to a piece of information that is spreading rapidly~(resp.~slowly) through the social network. 

% The question of when to use vanilla or the friendship paradox-based stochastic approximation should be applied in a given setting depends on statistical properties of the two estimates. % on which they are based (given in \cref{eq:VanillaEstimate} and \cref{eq:FPEstimate}) - 
% We address this question in detail in Sec.~\ref{sec:theoretical_comparison}~(theoretically), Sec.~\ref{sec:numerical_results}~(numerically) and Sec.~\ref{sec:empirical_results}~(empirically).

In summary, Sec.~\ref{sec:extensions} extended the vanilla and friendship paradox-based estimators proposed in Sec.~\ref{sec:algorithms} to two settings:~directed networks and dynamic information cascades. These extensions are numerically and empirically evaluated in the subsequent sections.

\section{Numerical Experiments}
\label{sec:numerical_results}

\begin{figure}
    \centering
    \begin{subfigure}{1\columnwidth}
    	\centering
        \includegraphics[width=\linewidth, trim=0.1in 0.1in 0.1in 0.3in, clip]{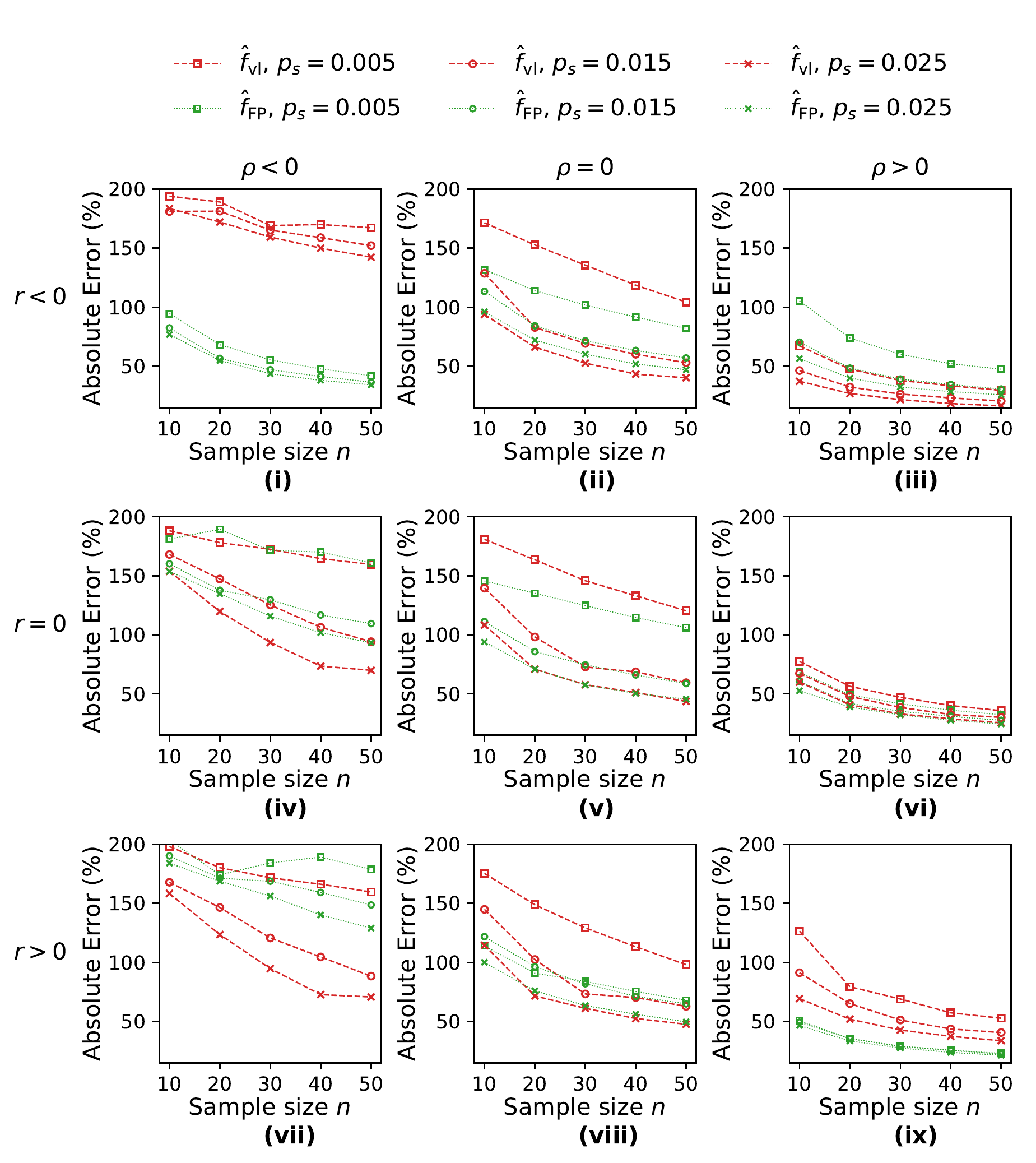}
        \caption{Absolute error values for a power-law exponent $\mathbf{\exponent = 2.5}$}
        \label{subfig:AbsError_Alpha2pt5}
    \end{subfigure}
    \begin{subfigure}{1\columnwidth}
        \centering
        \includegraphics[width=\linewidth, trim=0.1in 0.1in 0.1in 1.25in, clip]{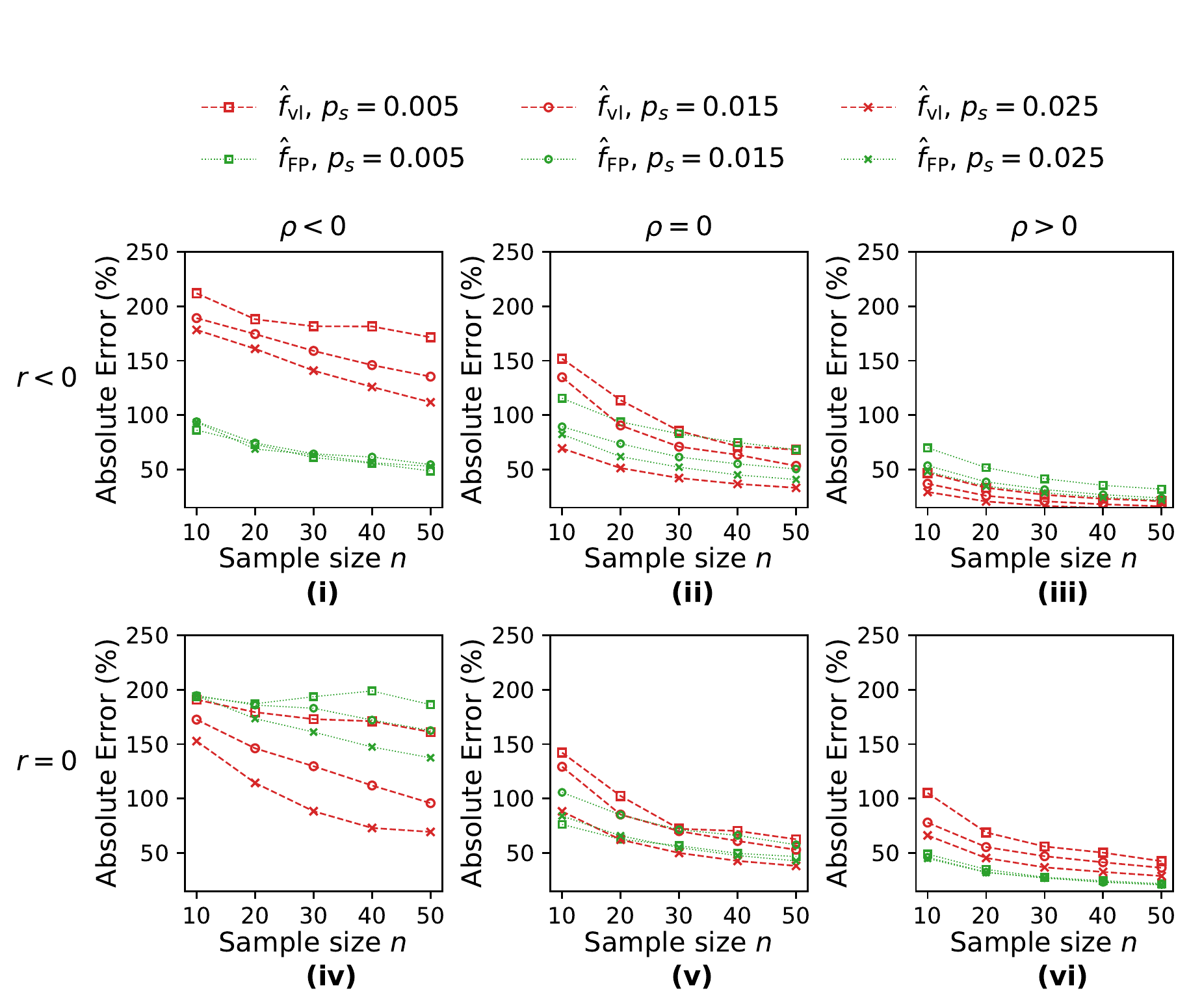}
            \caption{Absolute error values for a power-law exponent $\mathbf{\exponent = 2.2}$. 
            % Due to the heavier tail of the network with the power-law exponent $\mathbf{\exponent = 2.2}$ (compared to the network with $\mathbf{\exponent = 2.5}$), the edges of the network cannot be re-wired to make it assortative (i.e.,~$\AssortativityCoefficient>0$) using the simulation procedure described in Sec.~\ref{sec:numerical_results}. \textcolor{red}{This limitation is justified by the theoretical result in \cite{van2014degree} that $\DegreeSharingCorrelationCoefficient$ is non-negative for infinite heavy-tailed networks}. Hence, only $\AssortativityCoefficient<0$ (disassortative) and $\AssortativityCoefficient=0$ (neither disassortative nor assortative) are considered for this network. 
            }
        \label{subfig:AbsError_Alpha2pt2}
	\end{subfigure}
	\caption{The absolute error values of the vanilla estimate~$\VanillaEstimate$ (given in \cref{eq:VanillaEstimate}) and the friendship paradox-based estimate~$\FPEstimate$ (given in \cref{eq:FPEstimate}) for two synthetically generated power-law networks (with power-law exponents $\mathbf{\exponent = 2.5}$ and $\mathbf{\exponent = 2.5}$) with various values of the assortativity coefficient~$\mathbf{\AssortativityCoefficient}$, degree-sharing correlation coefficient~$\mathbf{\DegreeSharingCorrelationCoefficient}$ and the unconditional probability of sharing the piece of information $\unconditionalSharinProb$~(i.e.,~$\unconditionalSharinProb = \sum_k\conditionalSharingDist\left(1|k\right)\DegDist\left(k\right)$ is the probability that a uniformly chosen node shares the piece of information). The shown values were estimated via a Monte Carlo simulation as explained in Sec.~\ref{sec:numerical_results}. The plots show that the numerical results agree with the conclusions reached in the statistical analysis in Sec.~\ref{sec:theoretical_comparison}. 
	In particular, the $\FPEstimate$ is the better choice when both $\mathbf{\AssortativityCoefficient}$ and $\mathbf{\DegreeSharingCorrelationCoefficient}$ have the same sign. Further, comparing Fig~\ref{subfig:AbsError_Alpha2pt5} with Fig.~\ref{subfig:AbsError_Alpha2pt2} shows that heavy-tails increase the disparity between the performances of the two estimators.}
	\label{fig:AbsError_of_Estimates}
\end{figure}

% \begin{figure}[t!]
%     \centering
%     \includegraphics[width=\linewidth, trim=0.1in 0.1in 0.1in 0.1in, clip]{Figures/NEW_ICM_A_10k_alpha2pt5_rkk_0pt2.npy_InfectionProb_0.05_StepSize_0.01_Ratio_100}
%     \caption{The performance of the two stochastic approximation algorithms based on the vanilla and friendship paradox-based estimates~(given in \cref{eq:vanilla_SA} and \cref{eq:FP_SA}, respectively) for tracking the exposure to an information cascade under the Independent Cascade model~(ICM) for a synthetic power-law network with the power-law exponent $\mathbf{\exponent = 2.5}$. }
%     \label{fig:SA_AbsError_Alpha2pt5_ICM_rkk_positive}
% \end{figure}
	
\begin{figure*}[t!]
    \centering
    \begin{subfigure}[T]{0.485\textwidth}
    	\centering
        \includegraphics[width=\linewidth, trim=0.1in 0.1in 0.1in 0.1in, clip]{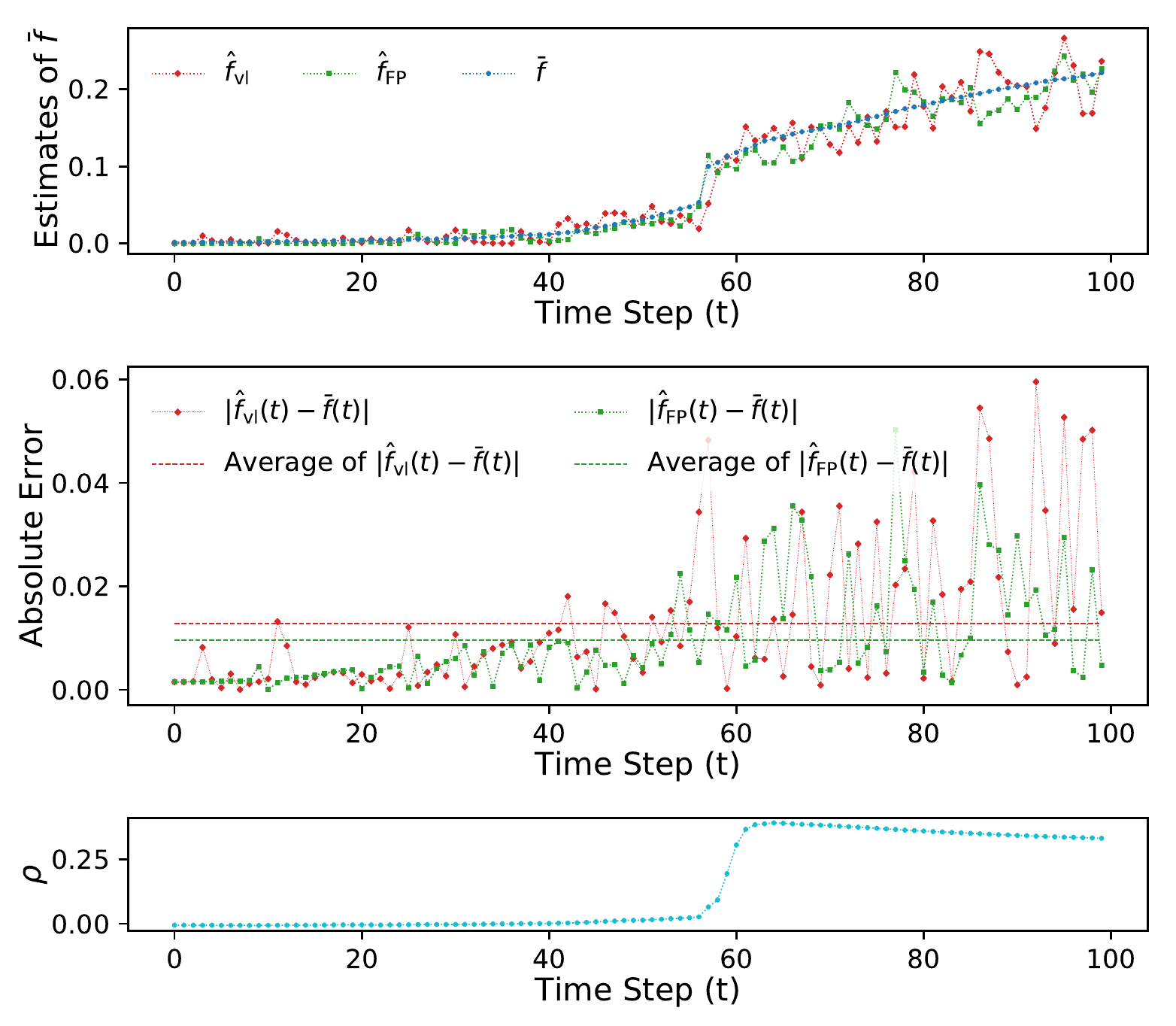}
        \caption{Assortative network (i.e.,~assortativity coefficient~$\AssortativityCoefficient>0$)}
        \label{subfig:SA_AbsError_Alpha2pt5_ICM_rkk_assortative}
    \end{subfigure} \hfill
    \begin{subfigure}[T]{0.485\textwidth}
        \centering
        \includegraphics[width=\linewidth, trim=0.1in 0.1in 0.1in 0.1in, clip]{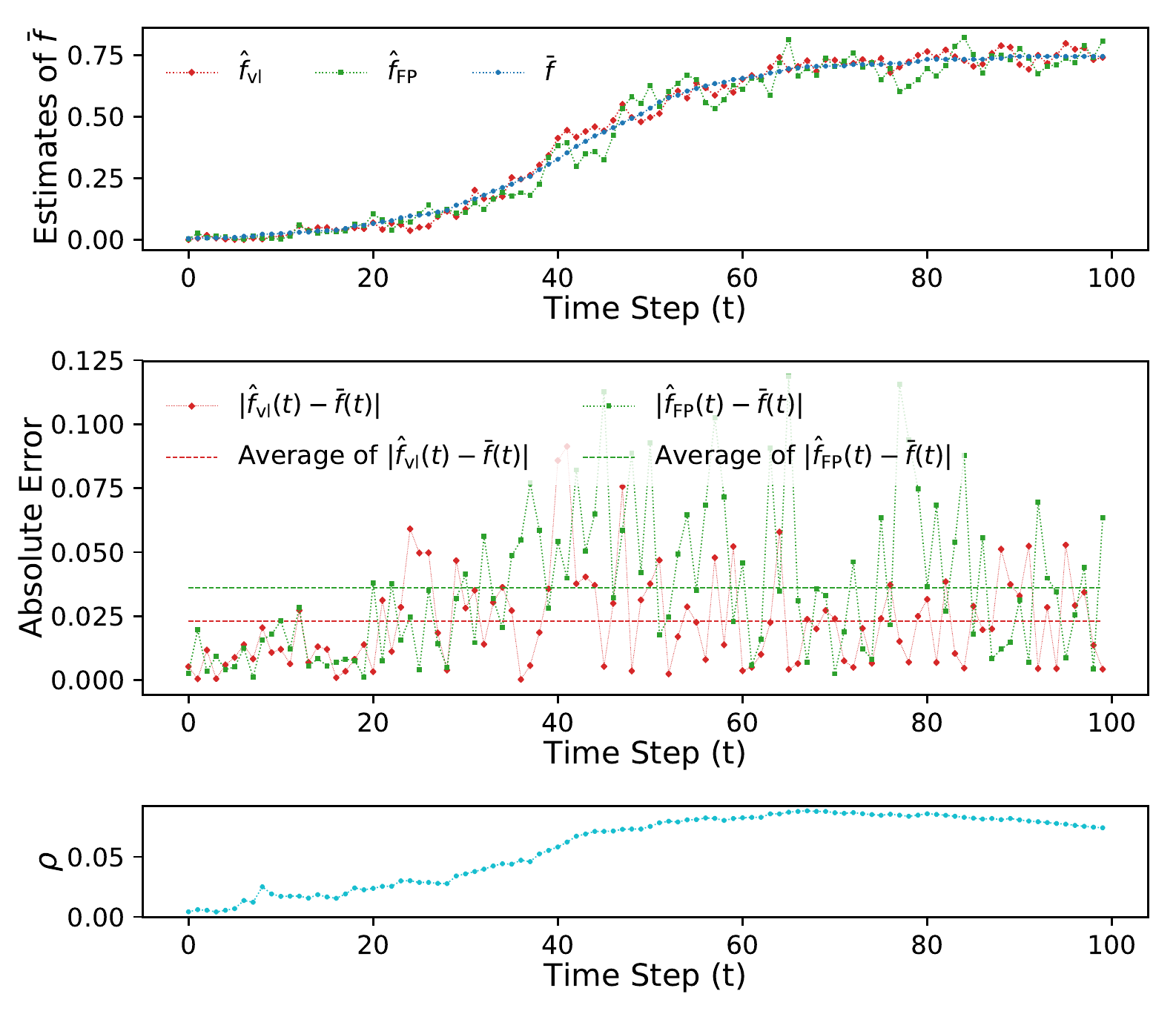}
        \caption{Disassortative network (i.e.,~assortativity coefficient~$\AssortativityCoefficient<0$)}
        \label{subfig:SA_AbsError_Alpha2pt5_ICM_rkk_disassortativ}
	\end{subfigure} \hfill
	\caption{The performance of the vanilla and friendship paradox-based stochastic approximation algorithms~(given in \cref{eq:vanilla_SA} and \cref{eq:FP_SA}, respectively) for tracking the exposure to an information cascade (in real-time) under the Independent Cascade model~(ICM) on a synthetic power-law network with the power-law exponent $\mathbf{\exponent = 2.5}$. The top-row shows the two estimates together with the true parameter $\trueparameter$ and the middle-row shows the absolute errors corresponding to the two estimates at each time instant and the average error over all time instants. The third row shows the variation of the degree-sharing correlation coefficient $\DegreeSharingCorrelationCoefficient$ with time. 
% 	For the shown results, it is assumed that $100$ samples are obtained at each time step (i.e.,~the diffusion process evolves on a 100 times slower time scale compared to obtaining the samples), the step-size $\stepsize = 0.01$ and, the adoption probability is $0.05$~(each neighbor of a person who shared the piece of information shares it independently with probability $0.05$ in the subsequent time instant.)
The friendship paradox-based stochastic approximation algorithm works better for the assortative network (Fig.~\ref{subfig:SA_AbsError_Alpha2pt5_ICM_rkk_assortative}) while the the vanilla stochastic approximation algorithm works better for the disassortative network~(Fig.~\ref{subfig:SA_AbsError_Alpha2pt5_ICM_rkk_disassortativ}). This observation agrees with the theoretical conclusions reached in Sec.~\ref{sec:theoretical_comparison}.}
	\label{fig:SA_AbsError_Alpha2pt5_ICM_rkk_positive_and_negative}
\end{figure*}	
	
In this section, we numerically compare the two estimators (vanilla estimator $\VanillaEstimate$ given in \cref{eq:VanillaEstimate} and the friendship paradox-based estimator $\FPEstimate$ given in \cref{eq:FPEstimate}) using detailed simulation experiments. The aim of this comparison is to verify and complement the theoretical analysis in Sec.~\ref{sec:theoretical_comparison}, and obtain additional insight into the performance of the two methods and how they compare with each other in various settings.

\vspace{0.2cm}
\noindent
{\bf Simulation setup: } To compare the vanilla estimator $\VanillaEstimate$ and the friendship paradox-based estimator $\FPEstimate$~(given in \cref{eq:VanillaEstimate} and \cref{eq:FPEstimate}), we use synthetically generated power-law networks from the configuration model~\cite{newman2003structure} with $10000$ nodes and power-law exponents $\exponent = 2.5$ and $\exponent=2.2$. The edges in each network $G = (V,E)$ are re-wired using the method proposed in \cite{van2010influence} to explore the three possible regions of the assortativity coefficient $\AssortativityCoefficient \in [0,1]$ ($\AssortativityCoefficient<0, \AssortativityCoefficient=0, \AssortativityCoefficient>0$). Next, the value $\shareInfo(v) \in \{0,1\}$ for each node $v\in V$ is first assigned as an iid Bernoulli random variable whose parameter determines the fraction of the people that share the piece of information. Then, the values $\shareInfo(v), v\in V$ are swapped amongst the nodes using the label-swapping method used in \cite{lerman2016majority} to correlate the sharing and degree to consider the three cases of the degree-label correlation coefficient~$\DegreeSharingCorrelationCoefficient \in [0,1]$~($\DegreeSharingCorrelationCoefficient<0$,  $\DegreeSharingCorrelationCoefficient=0$ and $\DegreeSharingCorrelationCoefficient>0$). The results obtained using this simulation setup are shown in Fig.~\ref{fig:AbsError_of_Estimates}. To compare the recursive algorithms based on the vanilla and friendship paradox-based estimators~(given in \cref{eq:vanilla_SA} and \cref{eq:FP_SA}, respectively), we use two well-known information diffusion models: the Independent Cascade model (ICM) and the Linear Threshold Model~(LTM). The results for the ICM are shown in Fig.~\ref{fig:SA_AbsError_Alpha2pt5_ICM_rkk_positive_and_negative} and the results for the LTM are given in Appendix~\ref{sec:appendix_additional_details_and_more_results}. 
% For the ICM, we consider the adoption probability to be $0.05$~(i.e.,~each friend of a person who has shared a piece of information shares it independently with probability $0.05$ in the following time steps). 
% For LTM, we assume a person shares a piece of information if $5\%$ of more of her neighborhood has shared the piece of information in the previous time instant. We set the step-size $\stepsize = 0.01$ and assume that the diffusion process evolves for each $100$ samples that we collect~(i.e.,~the diffusion evolves on a time scale that is $100$ times slower). 
% Additional details about the simulation setup are provided in the Appendix~\ref{sec:appendix_additional_details_and_more_results} and, all simulation codes are publicly available in the Github repository to ensure the full reproducibility of the results.\bn{Github repo not yet created.}
% \bnedit{All simulations codes with additional details will be made publicly available for full reproducibility.}
{Additional details about the simulation setup are given in Appendix~\ref{sec:appendix_additional_details_and_more_results}.}

\vspace{0.2cm}
\noindent
{\bf Discussion of the Numerical Results (Fig.~\ref{fig:AbsError_of_Estimates} and Fig.~\ref{fig:SA_AbsError_Alpha2pt5_ICM_rkk_positive_and_negative}): } The numerical results verify the theoretical conclusions (from Sec.~\ref{sec:theoretical_comparison}) and yield additional insight into the practical usefulness of the two methods as we discuss below.

{\it 1.~Choosing the method that is best for the context: } Fig.~\ref{subfig:AbsError_Alpha2pt5} shows that the friendship paradox-based estimate~$\FPEstimate$ is more accurate~(compared to the vanilla estimate $\VanillaEstimate$) when the assortativity coefficient $\AssortativityCoefficient$ and the degree-sharing correlation coefficient~$\DegreeSharingCorrelationCoefficient$ have the same signs~(i.e.,~Fig.~\ref{subfig:AbsError_Alpha2pt5}(i) and Fig.~\ref{subfig:AbsError_Alpha2pt5}(ix)). When the assortativity coefficient $\AssortativityCoefficient$ and the degree-sharing correlation coefficient~$\DegreeSharingCorrelationCoefficient$ have different signs~(i.e.,~Fig.~\ref{subfig:AbsError_Alpha2pt5}(iii) and Fig.~\ref{subfig:AbsError_Alpha2pt5}(vii)), the vanilla estimate $\VanillaEstimate$ is more accurate compared to the friendship paradox-based estimate $\FPEstimate$. In addition, the vanilla estimate $\VanillaEstimate$ has a smaller error when the degree and sharing are uncorrelated~(i.e.,~$\DegreeSharingCorrelationCoefficient = 0$ corresponding to middle column of Fig.~\ref{subfig:AbsError_Alpha2pt5} and Fig.~\ref{subfig:AbsError_Alpha2pt2}) for when the sharing probability is not too small~(so that both methods yield absolute errors smaller than $100\%$ of the true parameter $\trueparameter$). Further, each subfigure of Fig.~\ref{subfig:AbsError_Alpha2pt5} and Fig.~\ref{subfig:AbsError_Alpha2pt2} shows that the difference in the accuracy of the two estimates is larger when the unconditional sharing probability $\unconditionalSharinProb$~(i.e.,~$\unconditionalSharinProb = \sum_k\conditionalSharingDist\left(1|k\right)\DegDist\left(k\right)$) is smaller, highlighting that choosing the best method is crucial when the piece of information has been shared by only a smaller fraction of people. These numerical observations verify the theoretical expectations captured in the first and second points in the discussion related to Theorem~\ref{th:variance_comparison_condition} in Sec.~\ref{sec:theoretical_comparison}, and emphasizes the importance of the choice for less widely shared pieces of information, per the third point.

\vspace{0.1cm}
{\it 2.~Implications of the heavy-tails: }Comparing Fig.~\ref{subfig:AbsError_Alpha2pt5} with Fig.~\ref{subfig:AbsError_Alpha2pt2} indicates that the difference in the accuracy of the two methods is larger when the tail of the degree distribution is heavier. Since real-world social networks have been empirically shown to have heavy-tails, this observation highlights the importance of utilizing the theoretical insight to pick the best method for the given context. 

\vspace{0.1cm}
{\it 3.~Tracking the exposure to an information cascade in real-time: } Fig.~\ref{fig:SA_AbsError_Alpha2pt5_ICM_rkk_positive_and_negative} shows the performance of the vanilla and the friendship paradox-based stochastic approximation algorithms~(\cref{eq:vanilla_SA} and \cref{eq:FP_SA}, respectively) for tracking the exposure to an information cascade simulated from the ICM. In Fig.~\ref{subfig:SA_AbsError_Alpha2pt5_ICM_rkk_assortative}, it can be clearly seen that the friendship paradox-based stochastic approximation~\cref{eq:FP_SA} outperforms the vanilla stochastic approximation \cref{eq:vanilla_SA}, especially after the time-step $40$ where the diffusion process speeds up and the degree-sharing correlation coefficient starts increasing rapidly. In particular, the friendship paradox-based method closely detects the sudden phase transition of the diffusion process approximately at time-step $55$ where the exposure suddenly jumps to a larger value. This result aligns with the theoretical conclusions in Sec.~\ref{sec:theoretical_comparison}, since both assortativity coefficient $\AssortativityCoefficient$ and the degree-sharing correlation coefficient $\DegreeSharingCorrelationCoefficient$ are both positive, leading to the friendship paradox-based method to outperform the vanilla method. On the other hand, the Fig.~\ref{subfig:SA_AbsError_Alpha2pt5_ICM_rkk_disassortativ} corresponds to the disassortative networks. Since the assortativity coefficient $\AssortativityCoefficient$ and the degree-sharing correlation coefficient $\DegreeSharingCorrelationCoefficient$ have opposite signs, the vanilla method outperforms the friendship paradox-based method. 

\vspace{0.1cm}
{\it 4.~Implications of assortativity $\AssortativityCoefficient$ and degree-sharing correlation $\DegreeSharingCorrelationCoefficient$ on the overall accuracy: } 
It can be seen from Fig.~\ref{fig:AbsError_of_Estimates} that both estimates ($\VanillaEstimate$ and $\FPEstimate$) tend to be less accurate when the degree and sharing are negatively correlated~(i.e.,~$\DegreeSharingCorrelationCoefficient<0$ in first column of Fig.~\ref{fig:AbsError_of_Estimates}). Although the friendship paradox-based estimator performs better in Fig.~\ref{subfig:AbsError_Alpha2pt5}(i), its accuracy decreases when moving to Fig.~\ref{subfig:AbsError_Alpha2pt5}(iv,vii). This result is due to the fact that when $\DegreeSharingCorrelationCoefficient<0$, the nodes who share the piece of information are the less popular nodes, which makes the average exposure $\trueparameter$ smaller and more difficult to estimate. If $\AssortativityCoefficient<0$ in addition to $\DegreeSharingCorrelationCoefficient<0$~(e.g.,~a star graph where outer nodes are sharing), then the friendship paradox-based estimator $\FPEstimate$ can easily reach the core nodes that are more likely to be exposed due to their popularity, and thus reduce the variance as in Fig.~\ref{subfig:AbsError_Alpha2pt5}(i). However, when $\AssortativityCoefficient>0$ and $\DegreeSharingCorrelationCoefficient<0$~(i.e.,~Fig.~\ref{subfig:AbsError_Alpha2pt5}(vii)), the less popular fringe nodes who share the piece of information are more likely to be separated from the core of the network so that the friendship paradox-based estimator cannot reach them. As such, both estimators tend to be the least accurate when $\DegreeSharingCorrelationCoefficient<0, \AssortativityCoefficient>0$.
An important example of this is the case where a piece of information originates with the less visible (i.e.,~fringe) nodes of the network. As such, special attention should be paid to choosing the best method when $\DegreeSharingCorrelationCoefficient<0$ to get the best possible accuracy.

In summary, Sec.~\ref{sec:numerical_results} numerically
 compared the absolute errors of the estimates obtained using the two estimators presented in Sec.~\ref{sec:algorithms}~(vanilla estimator~$\VanillaEstimate$ and the friendship paradox-based estimator~$\FPEstimate$). The numerical results agree with the theoretical conclusions provided in Sec.~\ref{sec:theoretical_comparison} and shed more light on the conditions under which one method outperforms the other.

\section{Results on Real-World Networks}
\label{sec:empirical_results}

% In this section, we evaluate the proposed estimators using real-world network datasets to illustrate their practical feasibility and obtain additional insight.
{Evaluating the accuracy of the two estimators $\VanillaEstimate, \FPEstimate$ (proposed in Sec.~\ref{sec:algorithms}) requires the true exposure~$\trueparameter$~(i.e., the ground truth). However, as we stressed in Sec.~\ref{sec:intro}, the exact value of the ground truth~$\trueparameter$ depends on two features~(the full network and the set of sharers) which are highly difficult to obtain, and our study was motivated by this difficulty in the first place. As such, comparing the estimates with the ground truth $\trueparameter$ in real-world networks (e.g.,~Twitter, Facebook) is not feasible from a resource and computation viewpoint.}
% It is difficult to obtain complete real-world network data along with accurate sharing information.
Therefore, in this section, we first use real-world undirected networks with the sharing function generated synthetically~(Sec.~\ref{subsec:empirical_results_undirected}).
We then evaluate our estimators on a real-world network with actual sharing data, using the directed ACM citation network (Sec.~\ref{subsec:empirical_results_directed}). 

\subsection{Undirected Networks}
\label{subsec:empirical_results_undirected}
\begin{figure}
    \centering
    \begin{subfigure}{\columnwidth}
    	\centering
        \includegraphics[width=\linewidth, trim=0.1in 0.1in 0.1in 0.7in, clip]{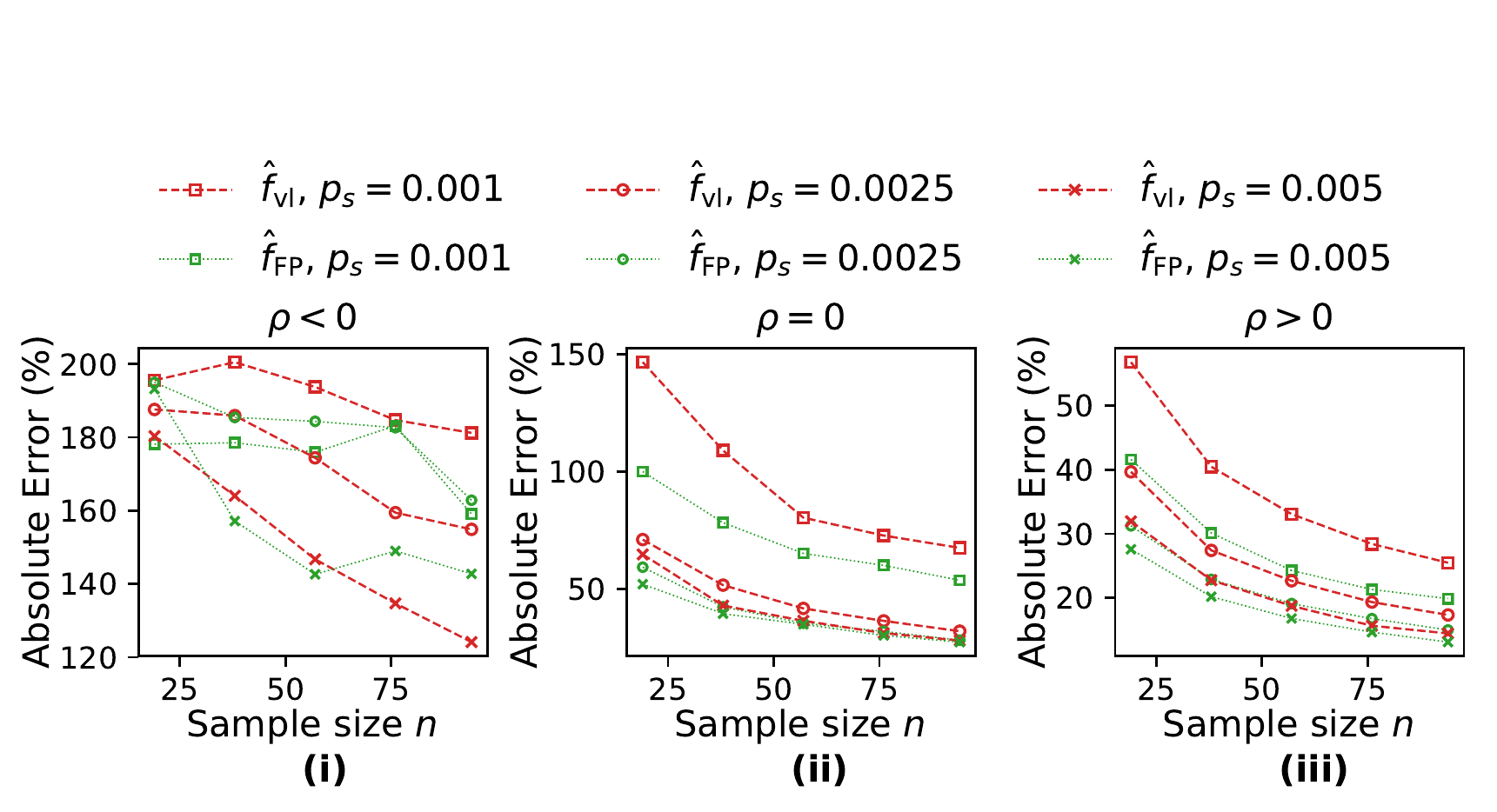}
        \caption{For a astrophysics co-authorship network ($\AssortativityCoefficient = 0.21$)}
        \label{subfig:AbsError_CA_AstroPh_v2_percent}
    \end{subfigure} 
    \begin{subfigure}{\columnwidth}
        \centering
        \includegraphics[width=\linewidth, trim=0.1in 0.1in 0.1in 1.3in, clip]{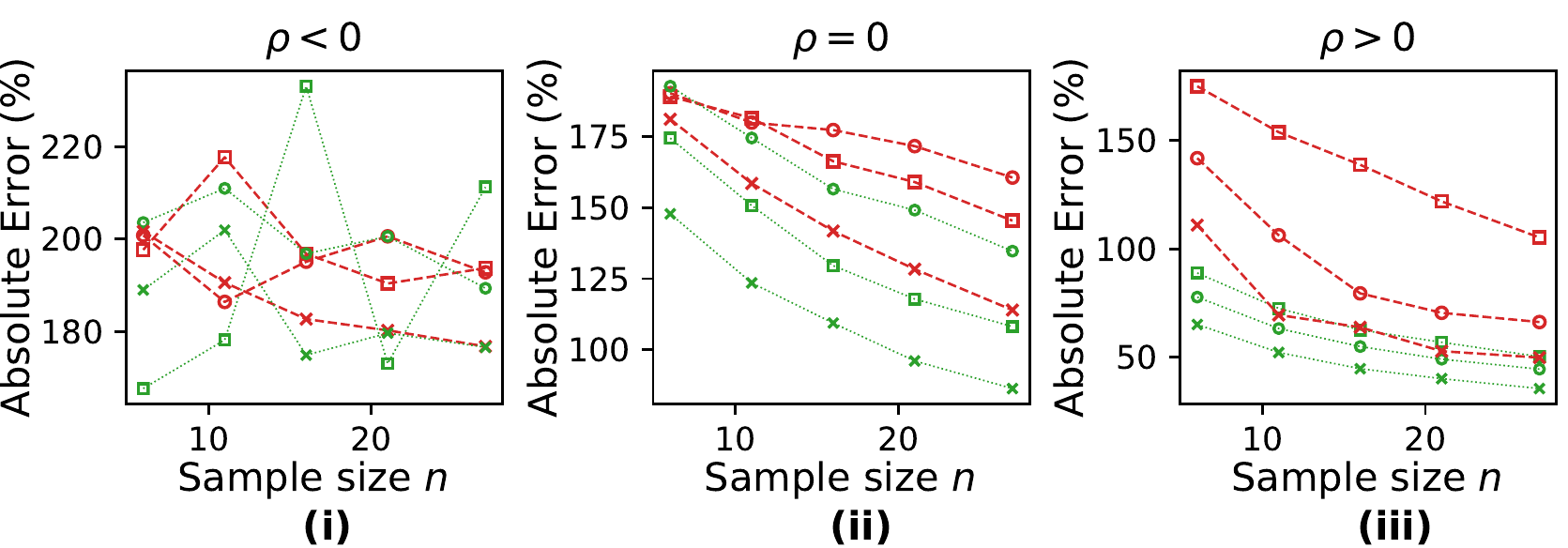}
            \caption{For a general relativity co-authorship network ($\AssortativityCoefficient = 0.66$)}
        \label{subfig:AbsError_CA_GrQc_v2_percent}
	\end{subfigure}
    \begin{subfigure}{\columnwidth}
        \centering
        \includegraphics[width=\linewidth, trim=0.1in 0.1in 0.1in 1.3in, clip]{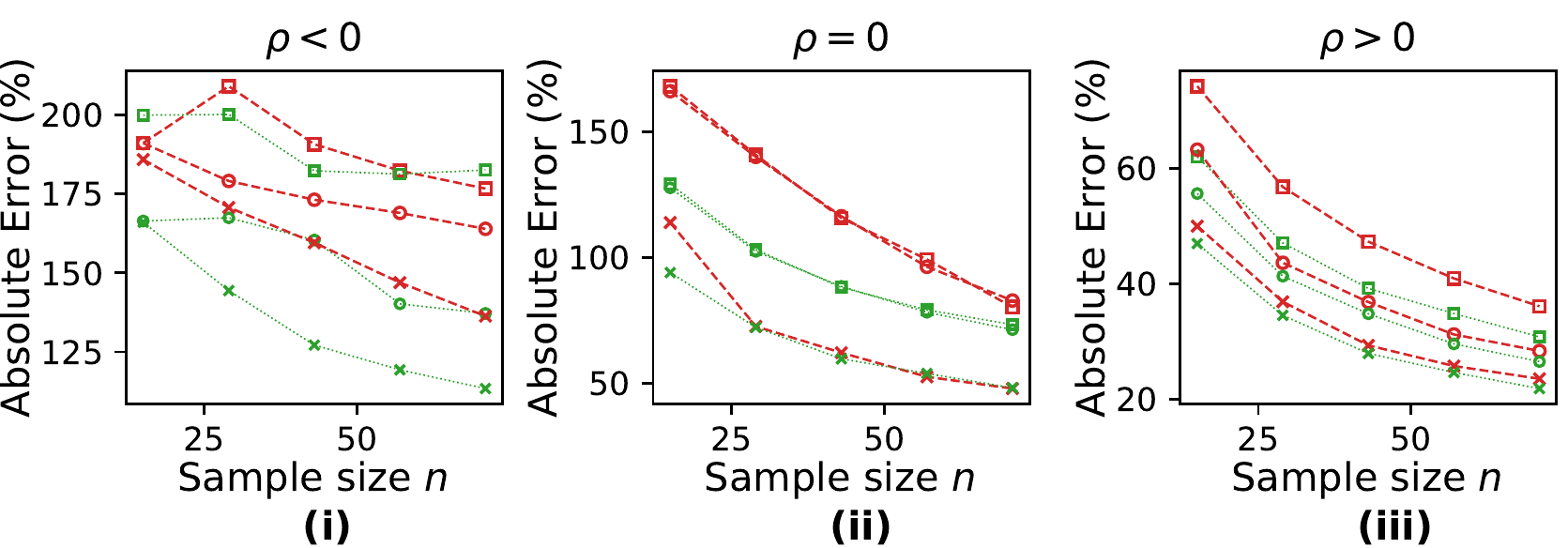}
            \caption{For a company Facebook page dataset ($\AssortativityCoefficient = 0.01$)}
        \label{subfig:AbsError_company_edges_v2_percent}
	\end{subfigure}	
    % \begin{subfigure}{\columnwidth}
%         \centering
%         \includegraphics[width=\linewidth, trim=0.1in 0.1in 0.1in 1.3in, clip]{Figures/NEW_AbsError_politician_edges_v2_percent.pdf}
%             \caption{Absolute error values for politician Facebook page dataset ($\AssortativityCoefficient = 0.02$)}
%         \label{subfig:AbsError_politician_edges_v2_percent}
% 	\end{subfigure}
	\begin{subfigure}{\columnwidth}
        \centering
        \includegraphics[width=\linewidth, trim=0.1in 0.1in 0.1in 1.3in, clip]{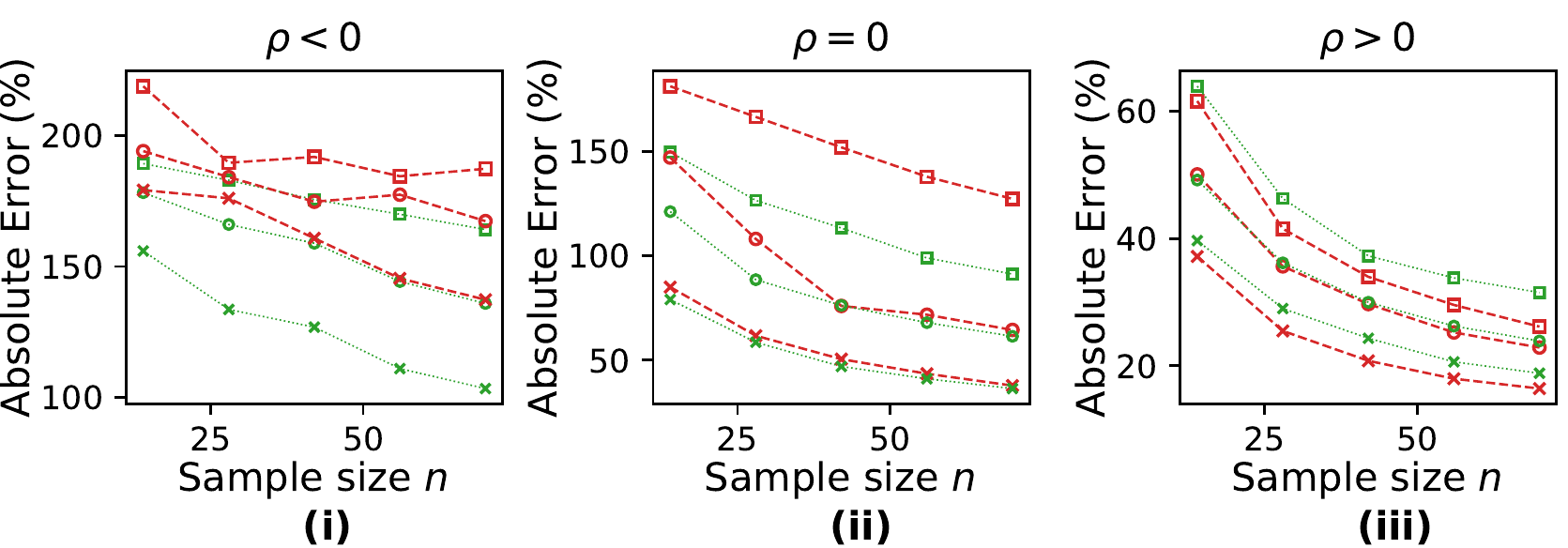}
            \caption{For an athlete Facebook page dataset ($\AssortativityCoefficient = -0.03$)}
        \label{subfig:AbsError_athletes_edges_v2_percent}
	\end{subfigure}\hfill 
	\caption{The absolute error values of the vanilla estimate~$\VanillaEstimate$ (given in \cref{eq:VanillaEstimate}) and the friendship paradox-based estimate~$\FPEstimate$ (given in \cref{eq:FPEstimate}) for four real-world network datasets. These results validate the theoretical insights (Sec.~\ref{sec:theoretical_comparison}) and complement the numerical experiments~(Sec.~\ref{sec:numerical_results}).}
	\label{fig:AbsError_of_Estimates_Empirical_Undirected}
\end{figure}

We tested the vanilla and friendship paradox-based estimators ($\VanillaEstimate, \FPEstimate$) on four publicly available real-world network datasets in the SNAP database~\cite{snapnets}. 
These networks include: a collaboration network between authors of papers submitted to Astrophysics and General Relativity in the Arxiv website, a network of Facebook pages of athletes,
% (where the nodes in the network represent the Facebook pages of athletes and the edges represent mutual likes among them)
and a Facebook page network of different companies.
% (where the nodes in the network represent the Facebook pages of companies and the edges represent mutual likes among them),
% and a network of Facebook pages of politicians.
% (where the nodes in the network represent the Facebook pages of politicians and the edges represent mutual likes among them). 
For these four networks, the sharing function $\shareInfo:V\rightarrow\{0,1\}$ was synthetically generated using the methods in Sec.~\ref{sec:numerical_results}. The results obtained using these five real-world networks are shown in Fig.~\ref{fig:AbsError_of_Estimates_Empirical_Undirected}.

For the network datasets corresponding to Fig.~\ref{subfig:AbsError_CA_AstroPh_v2_percent}-Fig.~\ref{subfig:AbsError_company_edges_v2_percent}~(where $\AssortativityCoefficient>0$), the friendship paradox-based estimator $\FPEstimate$ outperforms the vanilla estimator $\VanillaEstimate$ when $\DegreeSharingCorrelationCoefficient>0$ while both methods have relatively large error values~(above 100\%) when $\DegreeSharingCorrelationCoefficient<0$.
This result is as we expected since $\FPEstimate$ works better when $\AssortativityCoefficient, \DegreeSharingCorrelationCoefficient>0$ (as per the first point in the discussion related to Theorem~\ref{th:variance_comparison_condition}) and both $\VanillaEstimate, \FPEstimate$ tend to be less accurate when $\DegreeSharingCorrelationCoefficient<0, \AssortativityCoefficient>0$ (as per the fourth point in the discussion of numerical results). For the network dataset correponding to Fig.~\ref{subfig:AbsError_athletes_edges_v2_percent}, $\FPEstimate$~(resp.~$\VanillaEstimate$) works better when $\DegreeSharingCorrelationCoefficient<0$~(resp.~$\DegreeSharingCorrelationCoefficient>0$) since the network has $\AssortativityCoefficient<0$ (as we theoretically expected). Therefore, the empirical findings align with both the theoretical and numerical results~(Sec.~\ref{sec:theoretical_comparison} and Sec.~\ref{sec:numerical_results}, respectively).

\subsection{ACM Citation Network}
\label{subsec:empirical_results_directed}
In this analysis, we provide a 
full real-world network~~(i.e.,~without subsampling the network)
%complete
 as well as actual sharing data from that network. 
We use a network of academic papers and their references. 
We consider a paper that contains a phrase in its title as a ``sharer'' of that phrase and all papers that cite that paper as the ``exposed''.
We obtain a dataset of 629,814 papers from DBLP, ACM, and MAG (Microsoft Academic Group) \cite{citation_net}. 
We filtered out papers that did not have references within the original dataset or were not referenced by another paper in the original dataset to create a final dataset of 217,335 papers. 
We determine phrases of varying popularity to see how our estimates perform depending on the fraction of sharers. 
To generate the set of phrases, we first filter the papers' titles for stopwords and determine the frequency of each word to create a numerically sorted dictionary with word frequency pairs. 
Then, we use NLTK's bigram association measures to create word pairs (i.e., phrases) using a subset of words from the dictionary's beginning. 
We define popular phrases as having more than 400 sharers (e.g., \textit{data mining}, \textit{information systems}), average phrases with between 200 and 400 sharers (e.g., \textit{computer graphics}, \textit{embedded systems}), and unpopular phrases with 100 to 200 sharers (e.g., \textit{network design}, \textit{optimization problems}).
We perform this experiment with 25 popular phrases, 25 average phrases, and 25 unpopular phrases.

Based on the detailed results in Section~\ref{sec:theoretical_comparison} and their extension in Section \ref{subsec:directed_networks},
we expect the friendship paradox estimate based on random followers $\FollowerbasedEstimate$ to have the lowest absolute mean error, and the vanilla estimate $\VanillaEstimate$ to outperform the friendship paradox estimate based on random friends $\FriendbasedEstimate$. This is due to one version of the friendship paradox that states on average, a random follower has more friends than a random node; therefore, a random follower is more likely to be exposed to a piece of information. %\mn{because... (what is the condition -- i.e. what type of network we are seeing here -- just half a sentence, e.g. because this is a high-assortativity network... (or whatever). I know I said we don't need to explain, but we should just say what conditions we have that inform the expectation.} 
%\mn{I don't think we need to explain -- just point to the theoretical predictions above, right?}

\begin{figure}
\centering
        \centering
        \includegraphics[width=\linewidth]{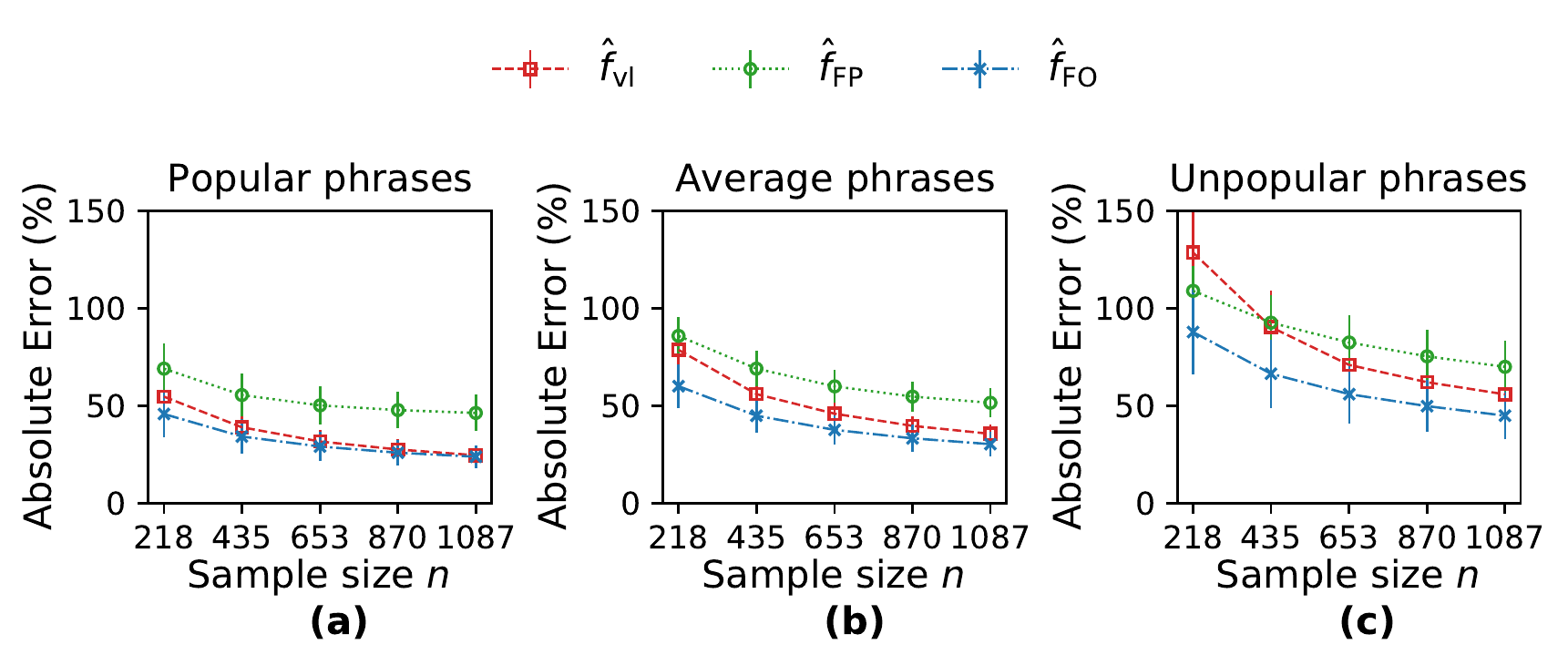}
    % \begin{subfigure}[T]{1\columnwidth}
    %     \centering
    %     \includegraphics [width=\linewidth]{Figures/TwitterResults2.pdf}
    % \caption{Absolute error values for the Twitter network}
    %  \label{subfig:twitterNetwork}
    % \end{subfigure}
\caption{The absolute error values of the vanilla estimate $\VanillaEstimate$, the friendship paradox estimate based on friends $\FriendbasedEstimate$, and the friendship paradox estimate based on followers $\FollowerbasedEstimate$ given in Eq.~(\ref{eq:DirectedEstimates}) on a real-world, directed network. These results validate the theoretical insights (Sec.~\ref{sec:theoretical_comparison}) and their extension (Sec. \ref{subsec:directed_networks}).}
\label{fig:empiricalResults}
\end{figure}

\vspace{0.2cm}
\noindent
{\bf Discussion of the ACM Citation Results (Fig.~\ref{fig:empiricalResults}):} 
Fig.~\ref{fig:empiricalResults} shows the average vanilla estimate $\VanillaEstimate$, friend-based estimate $\FriendbasedEstimate$, and follower-based estimate $\FollowerbasedEstimate$ for popular phrases (a), average phrases (b), and unpopular phrases (c). In each case, 
the follower-based estimate $\FollowerbasedEstimate$ outperforms the other two as we expected from the directed versions of the friendship paradox mentioned in Sec.~\ref{subsec:directed_networks}. In particular, Fig.~\ref{fig:empiricalResults} shows that the absolute error of all three estimates increase as the popularity of the phrases decrease. However, the follower-based estimate~$\FollowerbasedEstimate$ is more accurate compared to the other two estimates (i.e.,~the difference in the accuracy is larger) for unpopular phrases~(Fig.~\ref{fig:empiricalResults}(c)). This is because random followers are more likely to be exposed even to an unpopular piece of information due to their larger friend count~(according to the directed versions of the friendship paradox) and hence, sampling random friends lowers the variance of the estimate. In comparison, uniform and friend-based sampling are less likely to reach the smaller number of exposed individuals when the popularity of the piece of information is lower and as such, they yield a larger variance.
% the friendship paradox estimate based on followers $\FollowerbasedEstimate$, outperforms the vanilla estimate $\VanillaEstimate$ and the friendship paradox based on friends $\FriendbasedEstimate$, which agrees with the predictions from above. 
%\mn{Say more here: that it happens for all types of phrases; that it agrees with the predictions from above (or not). }

\section{Conclusion}
% This paper considered the problem of estimating exposure to information in social networks. 
% Although the mechanisms through which people are exposed to information in them have been well-studied, 
We presented 
a practically feasible framework for estimating the fraction of people who have been exposed to a piece of information by their contacts~(i.e.,~average exposure) in a social network.
% is lacking in the literature. 
% As a solution, 
In particular,
we proposed two methods to estimate the average exposure: a method that is based on uniform sampling and a method that samples random friends~(random ends of random links). The latter method can be thought of as a variance reduction method motivated by the friendship paradox which incorporates more high-degree individuals~(who are more likely to be exposed to the piece of information due to their large popularity) into the sample. Both methods are unbiased; we provided  theoretical results which characterize the conditions (in terms of properties of the underlying network as well as the piece of information) where one method outperforms the other
in terms of the variance. 
We also presented extensions of the proposed methods to directed networks and dynamic information cascades~(where the average exposure needs to be tracked in real-time). 
The proposed methods and the theoretical conclusions were verified numerically~(via simulations) as well as via experiments on several real-world network datasets.
% \begin{acks}
%To Robert, for the bagels and explaining CMYK and color spaces.
% \end{acks}

\bibliographystyle{ACM-Reference-Format}
\bibliography{EstimatingExposure_ref}

\newpage
\appendix

\section{Proofs of Theorems}
\subsection{Proof of Theorem~\ref{th:Bias_and_Variance}}
\label{subsec:proof_th_Bias_and_Variance}

\emph{Part~1:}~For the vanilla estimate $\VanillaEstimate$, it follows that $\expec\left\{\VanillaEstimate\right\} = \trueparameter$ since it is the average of $\numsamples$ iid Bernoulli random variables (with parameter $\trueparameter$). For the friendship paradox-based estimate $\FPEstimate$,
\begin{align*}
    \expec\left\{\FPEstimate\right\} &= \expec\left\{\frac{\avgdegree}{n}\sum_{i = 1}^{\numsamples}\frac{\exposureInfo(Y_i)}{\degree(Y_i)}\right\}\\
    &={\avgdegree}\expec\left\{\frac{\exposureInfo(Y_1)}{\degree(Y_1)}\right\} \quad \text{($\because$ $Y_1, \dots, Y_\numsamples$ are iid samples.)}\\
    &= \avgdegree \sum_{v \in V} \frac{\exposureInfo(v)}{\degree(v)} \times \frac{\degree(v)}{\sum_{v\in V}\degree(v)} \quad \text{($\because$ $\prob (Y_1 = v) = \frac{\degree(v)}{\sum_{v\in V}\degree(v)} $)}\\
    &= \avgdegree \sum_{v \in V} \frac{\exposureInfo(v)}{\numsamples\avgdegree} = \sum_{v \in V} \frac{\exposureInfo(v)}{\numsamples} =  \trueparameter.
\end{align*} Therefore, both $\VanillaEstimate, \FPEstimate$ are unbiased estimates of 
% the average exposure 
$\trueparameter$.

\emph{Part~2:}~Consider the variance of the vanilla estimate $\VanillaEstimate$. Since the estimate is the average of $\numsamples$ iid Bernoulli random variables (with parameter $\trueparameter$), their variance is given by $\trueparameter(1-\trueparameter)/\numsamples$. For the friendship paradox-based estimate $\FPEstimate$,
\begin{align*}
    \var\left\{\FPEstimate\right\} &=  \var\left\{\frac{\avgdegree}{n}\sum_{i = 1}^{\numsamples}\frac{\exposureInfo(Y_i)}{\degree(Y_i)}\right\} \\
    &= \frac{1}{n}\var\left\{\avgdegree\frac{\exposureInfo(Y_1)}{\degree(Y_1)}\right\}  \quad \text{($\because$ $Y_1, \dots, Y_\numsamples$ are iid samples.)}\\
    &= \frac{1}{n}\left(\expec\left\{\left(\avgdegree\frac{\exposureInfo(Y_1)}{\degree(Y_1)}\right)^2\right\} - \trueparameter^2 \right) \quad \text{($\because$ $\expec\left\{\FPEstimate\right\} = \trueparameter$)}\\
    &=\frac{1}{n}\left( \avgdegree^2\sum_{v\in V} \frac{\exposureInfo^2(v)}{\degree^2(v)} \times \frac{\degree(v)}{\sum_{v\in V}\degree(v)} - \trueparameter^2 \right) \\ & \quad \text{($\because$ $\prob (Y_1 = v) = \frac{\degree(v)}{\sum_{v\in V}\degree(v)} $)}\\
    &=\frac{1}{n}\left( \avgdegree^2\sum_{v\in V} \frac{\exposureInfo(v)}{\degree(v)} \times \frac{1}{\numsamples\avgdegree} - \trueparameter^2 \right) = \frac{1}{n}\left( \avgdegree\expec\left\{\frac{\exposureInfo(X)}{\degree(X)}\right\} - \trueparameter^2 \right).
\end{align*}

\subsection{Proof of Theorem~\ref{th:variance_comparison_condition}}
\label{subsec:proof_th_variance_comparison_condition}
Note that $\conditionalDegDist\left(k'|k\right)\conditionalSharingDist\left(0|k'\right)$ is the probability that a degree $k$ node connects to a degree $k'$ node that hasn't shared the piece of information. Therefore, averaging this term over the value $k'$ \\(i.e.,~$\sum_{k'}\conditionalDegDist\left(k'|k\right)\conditionalSharingDist\left(0|k'\right)$) yields the probability that a degree $k$ node having a neighbor that hasn't shared the piece of information. For a degree $k$ node to not be exposed to the information, all $k$ neighbors of that node must not have shared the piece of information. Hence, $\left(\sum_{k'}\conditionalDegDist\left(k'|k\right)\conditionalSharingDist\left(0|k'\right)\right)^k$ is the probability that a node with degree $k$ has not been exposed to the piece of information~i.e.,~
\begin{equation}
    \prob\left\{f(X) = 0|\degree(X) = k\right\} = \left(\sum_{k'}\conditionalDegDist\left(k'|k\right)\conditionalSharingDist\left(0|k'\right)\right)^k,
\end{equation}
which yields \cref{eq:prob_degree_k_exposed}.

Next, using the expressions for the variance in \cref{eq:variance} and the fact that $\trueparameter = \expec\left\{\exposureInfo(X)\right\}$ (where $X$ is a uniformly sampled node), we get,
\begin{align*}
     &\var\{\VanillaEstimate\} \geq \var\{\FPEstimate\}  \iff \frac{1}{\numsamples}\left( \avgdegree\expec\left\{\frac{\exposureInfo(X)}{\degree(X)}\right\} - \trueparameter^2\right) \geq \frac{1}{\numsamples}\trueparameter\left(1-\trueparameter\right) \nonumber \\
    &\hspace{0.2cm}\iff  \trueparameter - \avgdegree\expec\left\{\frac{\exposureInfo(X)}{\degree(X)}\right\} \geq 0 \nonumber \\
    &\hspace{0.2cm}\iff \expec\left\{\exposureInfo(X)\left(1 - \frac{\avgdegree}{\degree(X)}\right)\right\} \geq 0 \quad \text{($\because \expec\left\{\exposureInfo(X)\right\} = \trueparameter$ from \cref{eq:bias})} \nonumber \\
    &\hspace{0.2cm}\iff \mathbb{E}_{k \sim \DegDist}\left\{\expec\left\{\exposureInfo(X)\left(1 - \frac{\avgdegree}{\degree(X)}\right) \bigg|  \degree(X) = k\right\} \right\} \geq 0 \nonumber \\
    &\hspace{1cm}\text{(by conditioning on $\degree(X) = k$ and then averaging over $k$)} \nonumber \\
    &\hspace{0.2cm} \iff \mathbb{E}_{k \sim \DegDist} \left\{ \left(1 - \frac{\avgdegree}{k}\right) \prob\left\{f(X) = 1|\degree(X) = k\right\}  \right\} \geq 0 \nonumber
\end{align*}
This completes the proof.

\section{Additional Numerical Results and Details for Reproducibility}
\label{sec:appendix_additional_details_and_more_results}

% All simulations were performed using version 3.7.3 of the software Python. 
The Github repository will be made publicly available for full reproducibility.
Below, we provide additional details related to the simulation setup used to generate the numerical results in Sec.~\ref{sec:numerical_results}. 

\vspace{0.2cm}
\noindent
\emph{Detailed Simulation setup for comparing the estimators $\VanillaEstimate, \FPEstimate$ (Fig.~\ref{fig:AbsError_of_Estimates}): } To generate the power-law networks, a sequence of $10k$ random variables from a power-law distribution with the required power-law exponent $\exponent$ were generated and rounded up to the nearest integer. Then, the first number in the sequence was altered by a value of $1$ if needed to make sure that the sum of the numbers is even (to be valid sequence of degrees). Then, the configuration model (\textit{configuration\_model} function in the \textit{networkx} package) was used to generate the networks with the given degree sequence. 

To change the assortativity of the generated power-law networks, we first sample two edges from the network uniformly (without replacement) and rewire them to increase or decrease the assortativity coefficient $\AssortativityCoefficient$ from the initial value. The Lemma~1 of \cite{van2010influence}, which orders the three possible ways to rewire the two selected edges based on the resulting assortativity coefficient values, is used to get the maximum increase or decrease in the assortativity coefficient when rewiring. This process is repeated until the required assortativiy coefficient values~($\AssortativityCoefficient \in \{-0.2,0, 0.2\}$) are reached or the maximum number of iterations ($100k$) is reached. 

To change the degree-sharing correlation coefficient~$\DegreeSharingCorrelationCoefficient$, we follow the attribute swapping procedure used in \cite{lerman2016majority} as follows. We first pick a node $u$ from the set of nodes who shared the piece of information uniformly~(i.e.,~$\shareInfo(u)=1$) and another node $v$ from the set of the people who has not shared the piece of information uniformly~(i.e.,~$\shareInfo(v)=0$). Then, to increase degree-sharing correlation coefficient~$\DegreeSharingCorrelationCoefficient$, we swap $\shareInfo(v)$ and $\shareInfo(u)$ if $\degree(u) < \degree(v)$~(resp.~$\degree(u) > \degree(v)$). This process is continued until the required degree-sharing correlation coefficient values~($\DegreeSharingCorrelationCoefficient \in \{-0.2,0, 0.2\}$) are reached or the maximum number of iterations ($100k$) is reached.

\begin{figure*}[t!]
    \centering
	\begin{subfigure}[T]{0.485\textwidth}
        \centering
        \includegraphics[width=\linewidth, trim=0.1in 0.1in 0.1in 0.1in, clip]{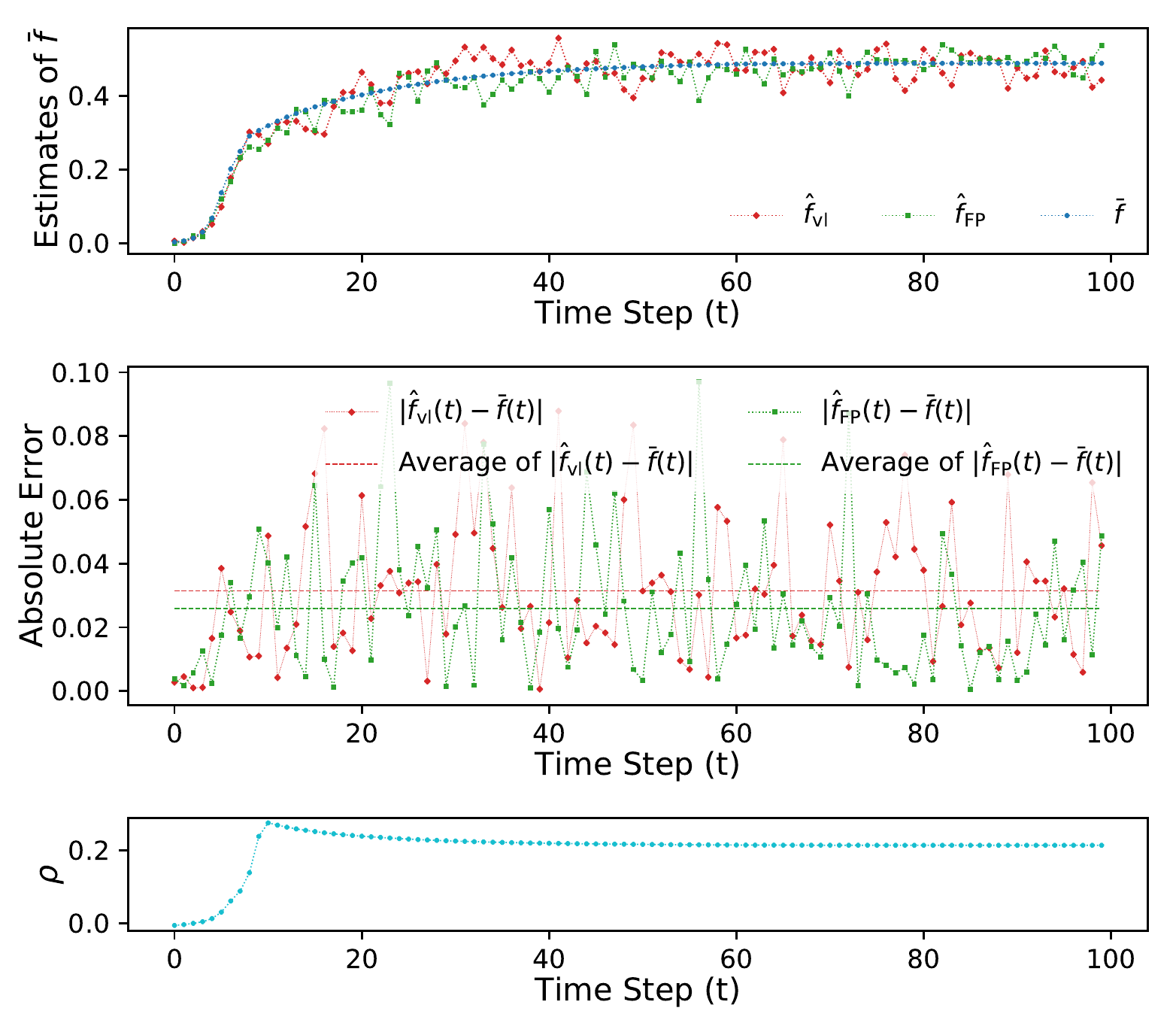}
        \caption{Assortative network (i.e.,~assortativity coefficient~$\AssortativityCoefficient>0$)}
	\end{subfigure} \hfill
	\begin{subfigure}[T]{0.485\textwidth}
        \centering
        \includegraphics[width=\linewidth, trim=0.1in 0.1in 0.1in 0.1in, clip]{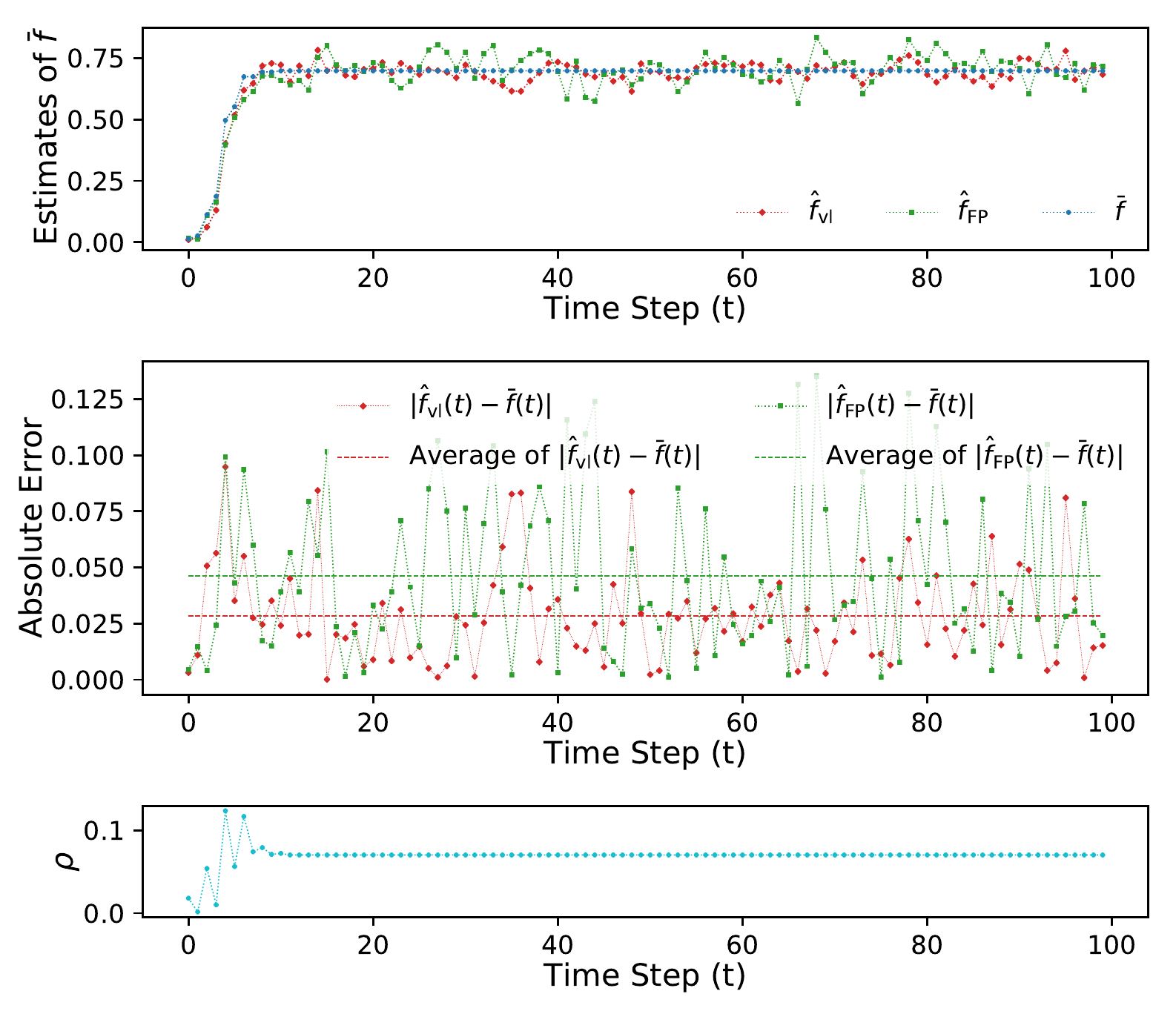}
        \caption{Disassortative network (i.e.,~assortativity coefficient~$\AssortativityCoefficient<0$)}
	\end{subfigure}
	\caption{The performance of the two stochastic approximation algorithms based on the vanilla and friendship paradox-based estimates~(given in \cref{eq:vanilla_SA} and \cref{eq:FP_SA}, respectively) for tracking the exposure to an information cascade under the Linear Threshold Model~(LTM) for a synthetic power-law network with the power-law exponent $\mathbf{\exponent = 2.5}$. This result complements the reult shown in Fig.~\ref{fig:SA_AbsError_Alpha2pt5_ICM_rkk_positive_and_negative} for the Independent Cascade Model~(ICM). It can be seen that the conclusions reached under the ICM also hold for the LTM as well. Thus, the proposed vanilla and friendship paradox-based stochastic approximation algorithms can be used to track (in real-time) the exposure to information cascades with various dynamical properties.}
	\label{fig:SA_AbsError_Alpha2pt5_LTM_rkk_positive_and_negative}
\end{figure*}

\vspace{0.2cm}
\noindent
\emph{Simulation setup for comparing the vanilla and the friendship paradox-based stochastic approximations (given in \cref{eq:vanilla_SA} and \cref{eq:FP_SA}): } Under the 
ICM
% Independent Cascade Model~(ICM) 
used to generate Fig.~\ref{fig:SA_AbsError_Alpha2pt5_ICM_rkk_positive_and_negative}, each neighbor of a node who shared a piece of information at a previous time instant shares in the current time instant with a pre-specified probability named the \emph{infection probability}. For Fig.~\ref{fig:SA_AbsError_Alpha2pt5_ICM_rkk_positive_and_negative}, the diffusion was initialized with $10$ uniformly chosen nodes and the infection probability is set to $0.05$. Additionally, the step size $\stepsize$ of the stochastic approximations \cref{eq:vanilla_SA}, \cref{eq:FP_SA} is set to $0.01$. Further, it is assumed that the stochastic approximations \cref{eq:vanilla_SA}, \cref{eq:FP_SA} are updated $100$ times for each step of the diffusion process~i.e.,~the samples are collected $100$ times faster than the evolution of the diffusion process. 
% Intuitively, this corresponds to the practical context where we can collect $100$ samples from the network 
% within each day~(or hour) 
% for every time window
% where the average exposure $\true$ evolves. 

Fig.~\ref{fig:SA_AbsError_Alpha2pt5_LTM_rkk_positive_and_negative} shows analogous results obtained using 
LTM
% another diffusion model called the Linear Threshold Model 
where, a node shares a piece of information at the current time instant if the fraction of its neighbors that have already shared it by the previous time instant exceeds a certain threshold. We choose the threshold value to be $5\%$ for the Fig.~\ref{fig:SA_AbsError_Alpha2pt5_LTM_rkk_positive_and_negative}. The step-size of both stochastic approximations \cref{eq:vanilla_SA}, \cref{eq:FP_SA} as well as the number of samples collected at each time instant are the same as the case for the ICM. 

\section{Practical Implementation Details}
\label{sec:appendix_extensions}

% \bn{other definitions of exposure and consequently, other variations of the problem definition (based on link weights, number of interactions, etc.) \\multivariate $\trueparameter$ (several pieces of information),\\ more applications etc.}

% \bn{I think we can also move this paragraph to the Appendix~D (which is now titled "practical considerations" since directed network stuff is now in the main text in Sec.~\ref{sec:extensions})}

\noindent
{\bf When edges cannot be sampled uniformly: }The implementation of the friendship paradox-based estimate $\FPEstimate$ given in \cref{eq:FPEstimate} requires the uniform sampling of links from the underlying social network to obtain random friends~$Y_1, Y_2, \dots, Y_\numsamples$. This 
sampling approach
is feasible in situations where links have unique IDs from a range of integers and the link corresponding to a given integer can be accessed. 
In settings where 
such uniform edge sampling is not possible~(e.g.,~a fully unknown social network), the friendship paradox-based estimate~$\FPEstimate$ can be implemented via the use of random walks,
%This is due to the fact that the 
since a stationary distribution of a random walk on an undirected, connected, non-bipartite graph samples nodes with probabilities proportional to their degrees~(page 298,~\cite{durrett2010_probability}). 
Hence, the random variables $Y_1, Y_2, \dots, Y_\numsamples$ could be replaced with samples from a sufficiently long random walk. 
Alternatively, one can also use a second version of the friendship paradox which states that ``uniformly sampled friend of a uniformly sampled node has more friends than a uniformly sampled node, on average"~\cite{cao2016}. Hence, taking $\numsamples$ uniformly sampled nodes and then taking one random friend of each of them would also be an alternative approach for friendship paradox-based sampling in undirected networks. 
% \mn{This is actually the approach we take in our practical implementation, right? Do we want to say that here?}
% \bnedit{Analogously, in directed networks, random followers~(resp.~friends) of random nodes can be used to implement the estimator $\DirecteNetwrkRandomFollower$~(resp.~$\DirecteNetwrkRandomFriend$) in order to avoid having to sample edges uniformly. We exploited the latter approach in our empirical experiments in Sec.~\ref{fig:empiricalResults} which uses Twitter data.}

\vspace{0.2cm}
\noindent
{\bf When the set of sharers is known: }The set of sharers $\SetOfSharers = \{v\in V: \shareInfo(v) = 1\}$ maybe publicly known in some contexts~(e.g.~Twitter users who shared a particular hashtag). In such cases, several improvements can be made to the proposed methods. 

First, $\SetOfSharers$ is typically an array that can be ordered (e.g.,~a set of unique Twitter handles, a set of integer node IDs, etc.). Thus, for each sampled node $v\in V$, calculating $\exposureInfo(v) \in \{0,1\}$ becomes equivalent to the problem of finding out whether two ordered arrays~(the node $v$'s neighbors $\neighborset(v)$ and the set of sharers $\SetOfSharers$) intersect or not.
% ~i.e.,
% \begin{equation}
%     \exposureInfo(v) = \begin{cases}
%                         1  &\neighborset(v) \cap \SetOfSharers \neq \emptyset  \\
%                         0 &\neighborset(v) \cap \SetOfSharers = \emptyset
%                     \end{cases}
% \end{equation}
% where, $\emptyset$ denotes the empty set. 
Hence, it is computationally easier to calculate $\exposureInfo(v)$ when the set of sharers $\SetOfSharers \subset V$ is known. 

Second, when $\SetOfSharers \subset V$ is known, the average degree of the sharers $\expec\{\degree(X)|\shareInfo(X) = 1\} = \frac{\sum_{v\in S}\degree(v)}{|\SetOfSharers|}$ (which is typically hard to estimate if the set $\SetOfSharers$ is small compared to $V$) can be calculated. Further, the average degree of the people who have not shared (i.e.,~ $\expec\{\degree(X)|\shareInfo(X) = 0\}$) can be estimated by sampling. Then, comparing the two values can be used as a heuristic estimate of the sign of the degree-sharing correlation coefficient $\DegreeSharingCorrelationCoefficient$.

% \noindent
% {\bf Other definitions of Exposure to Information: }In this work, our definition of exposure to information was based on being a neighbor of a person who shared the piece of information. This definition could be extended to other} 

% \textcolor{red}{\section*{To be done}}
% Twitter data: 
% \begin{enumerate}
% \item Get full networks of 10 hashtags -- everyone following the people who shared. 
% \item Roughly how long does it take to get 1000 followers?
% \item Stream a day of 1\% Twitter gardenhose (\url{https://developer.twitter.com/en/docs/twitter-api/tweets/volume-streams/introduction}), keep user IDs for 1/1000 of tweets 
% \item Pick a number that we can handle based on the time estimate (e.g. 10,000), get the followers/friends for these random users set R. 
% \item Pick a friend/follower at random for each random user in R (Friend(R) and Follower(R) sets).
% \item Get the friends of users in Friend(R), Follower(R) (no need for followers).
% \item For each hashtag: ground truth (based on set overlap), baseline, and FP estimate based on the people in Follower(R), Friend(R) (use the same sets for all hashtags. 
% \end{enumerate}

% Maybe:
% \begin{enumerate}
% \item [Maybe] Add another ``naive method" based on the number of sharers and the average degree
% \end{enumerate}

% \textcolor{red}{\section*{Done}}
% \begin{enumerate}
%     \item Added Eq. to equation numbers in the text.
    
%     \item Removed the organization paragraph.
    
%     \item Removed ``Remark" environment
    
%     \item Shortened the subsection on the dynamic case
% \end{enumerate}

\end{document}